\documentclass{JHEP3}
\usepackage{amsmath, graphicx, amssymb, cite, afterpage}
\preprint{IPPP/05/11\\DCPT/05/22}
\author{S. A. Abel and James Gray\\
        Institute for Particle Physics Phenomenology and Department of 
        Mathematical Sciences, University of Durham, Durham, DH1 3LE, UK\\
        E-mail: \email{s.a.abel@durham.ac.uk}, \email{j.a.gray2@durham.ac.uk}}
\title{On the chaos of D-brane phase transitions}
\abstract{We study small instanton (and brane recombination) phase
transitions in phenomenological models built with D-branes. By explicitly
describing the cosmological dynamics of the moduli and matter fields,
we show that these transitions do not occur smoothly, but are typically
chaotic with the gauge group of the low energy theory fluctuating
in time. We comment on the potential implications for cosmological
questions such as inflation.}
\begin{document}

\section{Introduction: phase transitions on D-branes}

There are good reasons to believe that there was a {}``pre-thermal
epoch'' of the Universe in which phase transitions occurred that
were driven by the movement of moduli fields along flat or almost
flat directions. In this epoch gauge symmetries may have been restored
instead of broken, and there is no \emph{a priori} reason why symmetries
could not have been broken and restored more than once. If this happened
in the early Universe it would have far reaching implications for
such issues as inflation, baryogenesis and dark matter. It is of great
interest therefore to examine cases where moduli driven phase transitions
can be analysed in more detail than usual, especially if these cases
display such anomalous behaviour. 

The transitions we will discuss in this paper are of the so-called
\emph{small instanton} type \cite{Witten:1995gx}. They are known
to occur in both M-theory \cite{Ganor:1996mu} and D-brane 
models \cite{Douglas:1995bn}, but remarkably in the latter they can be analysed completely
in the small coupling and large volume limits in terms of the effective
field theory. This is largely due to the fact that the extra light
degrees of freedom which appear during such transitions are ``particles''
rather than tensionless non-critical strings as would be the case
in the M-theory equivalent. There are two complementary pictures of
what happens during a small instanton phase transition in type II
models: the more straightforward involves a low (e.g. 3) dimensional
D-brane impacting and dissolving on a stack of higher (e.g. 7) dimensional
ones; the less straightforward picture is the T-dual process, in which
for example two intersecting D6-branes undergo recombination. 
The brane recombination picture
provides a nice geometric realisation of symmetry breaking, but either
way the transition, when it evolves in this direction, is to a theory
with a gauge symmetry of lower rank. 
The issues we raise are therefore relevant to phase transitions in
the wide class of Standard Model (SM)-like models built from intersecting
or coincident D-branes 
\cite{branepheno,blumreview,branepheno37,cvetic:2001nr,branesm}. 
Indeed it may even be possible to find the SM itself,
with electroweak symmetry breaking realised in just such a manner
\cite{branesm}. 

In this work we examine the properties of 
these transitions when they take place in a cosmological setting. At first
sight it seems reasonable to suppose (and it has usually been assumed)
that such transitions would occur smoothly, with the lower dimensional
D-brane dissolving {}``monotonically'' in the higher dimensional
one. However we uncover some quite unexpected properties. The reason
behind this is that the effective action describing the transition is
related to one of the simplest chaotic systems. 
This leads to rather
curious physical phenomena, in which for example the low dimension
brane dissolves temporarily in the high dimension one, only to be
ejected at some later time. 
In addition
there is an attractor mechanism which means that the lower dimensional
D-brane never leaves the vicinity of the higher dimensional one for
very long. However it cannot ever be said to have \emph{fully} dissolved
in it either; the dissolved brane never completely spreads out, and
in fact this would require some additional ingredient such as supersymmetry
breaking. 
(The picture we describe here is valid at
scales significantly above the supersymmetry breaking scales suggested
by phenomenology and also at small coupling and low velocities.)
One additional feature which is of some interest is that the potential
has a term that is quartic in the instanton radius. The instanton
size is therefore constricted by the compact volume, an additional
factor working against the completion of the phase transition.

\subsection*{\textmd{\emph{The set-up }}}

The setting for our discussion is supersymmetric orientifold compactifications
of type II models involving various D-brane configurations \cite{orientifold}.
We concentrate on ${\cal{N}}=1$ supersymmetric compactifications with non-abelian
gauge fields and chiral fermions, which can be built with a gauge
group and particle spectrum that is fairly close to the MSSM \cite{branepheno,cvetic:2001nr}. 

Type IIB string theories include D7 and D3-branes. In the vacua we
shall be discussing D3-branes are present and are taken to span the
four dimensions of normal spacetime to preserve the four dimensional Poincar\'e
invariance of the vacuum. Various stacks of D7-branes are also present
and are taken to wrap our four dimensions for the same reason and
then each stack must wrap a compact 4 cycle in the orientifold. The
number of each type of brane wrapping the various possible cycles
is restricted by a set of tadpole cancellation conditions which essentially
implement the requirement that the net charge with respect to any
form field on a compact manifold is zero. Non-abelian gauge fields
and chiral fermions arise in these models on various brane world volumes
and intersection loci.

These type IIB constructions are related to other set-ups under various
dualities and it is worth briefly reviewing them as for certain situations
the alternatives can be more intuitive. First they are mirror to type
IIA orientifold constructions including configurations of intersecting
D6-branes in the vacuum. Once again the D6-branes are extended in
the four directions of our space and time. Their remaining dimensions
then wrap compact 3 cycles within the orientifold. In this case non-abelian
gauge fields arise on the D6-brane stack world volumes and chiral
fermions appear as extra light degrees of freedom where the various
D6-brane stacks have four dimensional intersections. Since this type
IIA set-up contains purely D6-branes, the only fields that are excited
in the background are the dilaton, the Ramond-Ramond one form and
of course the metric. In the S-dual picture in terms of eleven dimensional
M-theory these quantities all map to various parts of the metric.
Therefore the S-dual of these IIA constructions are purely metric
${\cal{N}}=1$ supersymmetric compactifications of eleven dimensional M-theory
to four dimensions. In other words these are compactifications of
M-theory on manifolds of $G_{2}$ holonomy. The presence of chiral
fermions and non-abelian gauge fields in the four dimensional effective
theory tells us that these $G_{2}$ compactifications must have various
singular surfaces within them. In particular, within the compact space,
codimension four singularities exist on which live the non-abelian
gauge fields \cite{Acharya:1998pm,Acharya:2000gb}. These intersect
at codimension 7 points and it is on these singularities that the
chiral fermions arise as extra light states \cite{Acharya:2001gy,Atiyah:2001qf,Witten:2001uq}.

Let us now return to the type IIB picture. Within this setup the small
instanton phase transitions occurs as follows: a D3-brane approaches a D7
brane stack, impinges upon it and dissolves to become an object comprised
primarily of D7 gauge fields \cite{Polchinski:1998rr}; the {}``fundamental''
D3-brane disappears and is replaced by a solitonic object, which is
essentially a Yang Mills instanton dressed up with the various other
fields that exist upon the D7-brane stack; at
the moment of dissolution the instanton appears as a point like 
object in the four dimensions orthogonal to the original D3-brane 
in the stack's world volume, but it is then free to spread out to 
become a solitonic three brane of finite width. 
From the four dimensional perspective this
process looks like a Higgs effect. The easiest way to see this is
to consider the four dimensional gauge symmetry which is associated
with the D7-brane stack. For a stack of $N$ branes away from any
special points within the orientifold for example, and in the absence
of any expectation values for the fields living on the D7-branes,
this is simply $U(N)$. As the D3-brane dissolves onto the the D7-brane
stack during the small instanton transition it gives rise to a non-vanishing
vacuum expectation value for the gauge fields associated with an $SU(2)$
subgroup of this $U(N)$. This is simply because the the Yang Mills
instanton on which the emerging solitonic object is based is valued
within $SU(2)$. As the relevant gauge field components have their
indices lying within the compact space this corresponds, in the four
dimensional effective theory, to giving $SU(2)$ valued expectation
values to Higgs fields.

This phenomenon will have duals in the IIA and M-theoretic
pictures. In IIA intersecting D6-brane models this process corresponds
to the collision of various D-brane stacks at four dimensional intersections
and their resulting recombination via some {}``fuzzy'' process 
\cite{cvetic:2001nr,Erdmenger:2003kn}.
\textbf{}Naturally the equivalent of this process in $G_{2}$ compactifications
is the recombination of codimension 4 singularities after collision
at a codimension 7 point. We should add that in the IIB picture we
will neglect fluxes on the D7 world volumes. In the IIA picture this
corresponds to the intersection angles being very small or zero. Extension
to the more realistic cases of finite intersection angles will be
considered in a future work. 

We now proceed to examine the dynamics of these processes. We present
the discussion in the type IIB language as this is the easiest in
which to phrase our methodology. However our results may be made applicable
to the other cases simply by taking the relevant mirror and then S
dualities.

The rest of this paper is structured as follows. In section 2 we briefly
review the relevant aspects of the realisation of the ADHM description
of instantons in D-brane physics. In section 3 we describe how this
becomes embedded in string phenomenological models. Section 4 solves
the resulting system to describe the dynamics of small instanton transitions
and discusses the late time behaviour of our system, and its implications
for cosmology.

\section{Instantons in D-brane physics}

We begin by recalling how the ADHM instanton construction of self
dual Yang Mills configurations \cite{Dorey:2002ik}, 
specifically single Yang Mills instantons,
is realised in D-brane physics \cite{Polchinski:1998rr}. This
will be crucial in understanding the physical interpretation
of the description of the small instanton transition process later
on. We shall start with the realisation of the ADHM formalism for
uncompactified D-brane world volumes, highlighting the ${\cal{N}}=2$ structure,
and then describe how this is modified when some of the branes are
wrapped upon compact cycles. Phenomenological models are ${\cal{N}}=1$ supersymmetric
and so have to be presented in suitable terms, thus in the next section
we will have to translate this into the ${\cal{N}}=1$ language, add in moduli
dependence and so forth. 

The realisation of the ADHM formalism in a D-brane context has been
understood for a number of years \cite{Polchinski:1998rr}. The system
we shall concentrate on comprises a D3-brane lying within a stack
of $N$ D7-branes in type IIB. As described above,
we shall take six of the ten dimensions
of space and time to be compactified on some appropriate compact space.
The D3-branes are taken to be oriented transversely to this compact
space and the D7-brane stack will be taken to wrap a four dimensional
submanifold within it. 

Now let us examine the effective four dimensional theory associated
with this set-up. First consider the limit in which the volume of
the 4 cycle which the D7-branes wrap is infinite; the world volume
of the D7-brane is then also infinite, and consequently
the gauge coupling of the D7-brane gauge symmetry 
(which is inversely proportional to the volume of the D7s) is zero.
In this limit therefore there is
no contribution to the potential of the $d=4$ system, which we take
to be ${\cal{N}}=2$ supersymmetric for now, from the $D$-term associated with
the D7 stack. We shall denote the ${\cal{N}}=2$ hypermultiplets 
arising from strings stretching
between the D3-brane and D7-brane stack as $\chi_{ai}$, where $i$
is the ${\cal{N}}=2$ $SU(2)$ automorphism index and $a$ is the $U(N)$
gauge index. (We will frequently refer to these states as `twisted'
because they are associated with non-integer mode expansions.) In
addition we shall write the scalar fields corresponding to the collective
coordinates of the D3-brane and the D7 stack in the two codimensions
of the D7-branes as $X^{I}$ and $X^{'J}$ respectively. The scalar
potential for the system in this ``infinite D7 volume limit'' is then
given by

\begin{equation}
V_{\textnormal{{Vol}}(T^{4})\rightarrow\infty}=\frac{g_{D3}^{2}}{4}\sum_{A=1}^{3}(\chi_{ia}^{\dagger}\sigma_{ij}^{A}\chi_{aj})^{2}+\sum_{I=1}^{2}\frac{(X^{I}-X^{'I})^{2}}{2\pi\alpha'}\chi^{\dagger}\chi\end{equation}
where $\sigma^{A}$ are the Pauli matrices. We note for later that
the gauge coupling $g_{D_{3}}$ is a function of the dilaton and moduli
fields. At the moment we assume that these are frozen, but in the
next section we shall have to include these as full dynamical degrees
of freedom. 

The potential has two branches of zeros. The first of these, the Coulomb
branch, corresponds to setting $\chi=0$ and then allowing the separation
of the branes to be arbitrary. This just corresponds to the D3-brane
and the D7 stack being free to move about in the transverse space
while not interacting with each another. The second branch of zeros,
the Higgs branch, is given by setting the separation of the D-branes
to zero so that the the D3-brane lies on top of the D7 stack and then
demanding that the D-flatness condition associated with the D3-brane
world volume gauge theory is satisfied,

\begin{equation}
\chi_{ia}^{\dagger}\sigma_{ij}^{A}\chi_{aj}=0.\end{equation}
It turns out that there is a precise connection between these D-flatness
conditions and the equations describing the collective coordinates
of localised instanton solutions on the D7-brane stack. Indeed the
above equation is none other than the ADHM constraint equations
in the canonical form for a solution with a single unit of instanton
charge in a $U(N)$ gauge theory, where $\chi$ is interpreted as
the ADHM data. The fields $\chi$ also have the $U(1)$ symmetry required
to complete the ADHM hyperK\"ahler quotient: this is simply the $U(1)$
of the gauge theory on the D3 world volume. The general set of solutions
to the D3 flatness condition of the Higgs branch is then given by
the following:
\begin{equation}
\chi_{ai}=\rho U_{ab}m_{bi}.\label{inst}\end{equation}
Here $\rho$ can be interpreted as \emph{the size of the instanton},
$U$ is an $\frac{SU(N)}{SU(N-2)\times U(1)}$ matrix containing the gauge orientation and embedding
moduli of the instanton within the gauge group and $m_{bi}=\delta_{bi}$.
The bottom line is that one can interpret the expectation values of
the twisted sector fields as describing the collective coordinates
of a Yang-Mills instanton configuration living in the D7 world volume,
and that this branch of solutions corresponds to a situation where
the D3-brane has dissolved upon the D7 stack. The degrees of freedom
associated with the Higgs branch are then just the collective coordinates
of the soliton that the D3-brane has become, which are in one to one
correspondence with the degrees of freedom of the Yang Mills instanton
on which the soliton is based (The associated instanton solution itself
can be detected by probe calculations if needs be \cite{Polchinski:1998rr}). 

\bigskip{}
This then is the situation when the world volume of the D7 stack is
taken to be infinite. How are things modified when we allow the D7
branes to wrap a submanifold of finite size? In this case there is
an additional contribution to the potential of the four dimensional
theory. This contribution comes from the D-terms associated with the
D7 world volume gauge theory, the gauge coupling of which is now finite.
The potential is now 

\begin{equation}
V=\frac{g_{D7}^{2}}{4}\sum_{A=1}^{3}\sum_{a=1}^{N^{2}}(\chi_{ic}^{\dagger}\sigma_{ij}^{A}t_{cd}^{(a)}\chi_{jd})^{2}+\frac{g_{D3}^{2}}{4}\sum_{A=1}^{3}(\chi_{ia}^{\dagger}\sigma_{ij}^{A}\chi_{ja})^{2}+\sum_{I=1}^{2}\frac{(X^{I}-X^{'I})^{2}}{2\pi\alpha'}\chi^{\dagger}\chi+\ldots\end{equation}
where the ellipsis indicates terms in the potential such as $D$-terms
involving the $X^{I}$ fields that are
zero in the interesting directions in field space. Here the $t$'s
are the generators of the $U(N)$ gauge group. Again the couplings
$g_{D_{3}}$ and $g_{D_{7}}$ will be \emph{different} functions of
the moduli fields describing the compact space.

Evidently the Coulomb branch of solutions is unaffected by the finiteness
of the compact space: the D3 and D7-branes can still move about unhindered
in the transverse space if the twisted sector is not excited. The
Higgs branch though is affected by the presence of the new term. In
this branch the potential reduces to the first two terms above. As
they are both positive definite for a zero of the potential they must
vanish independently at the global minimum. 
The vanishing of the second term simply implies
the D3 $D$-flatness condition as before. Thus any putative flat direction
that remains when the D7 volume is finite 
would then be a subspace of the instanton moduli space. The
question is if there is indeed 
a subspace of the instanton moduli
space that remains flat even in the presence of the extra potential
terms. The answer is no: substituting a general solution to the D3
$D$-flatness conditions into the first term in the potential we obtain
a non-zero contribution which is proportional to
the instanton size, $\rho$, to the fourth power. (We shall see
this explicitly in the next section.) Consequently the expectation value that
would be interpreted as the instanton size is no longer a flat direction
in the theory's moduli space; the dissolution of the D3-brane onto
the the D7 stack is no longer associated with a flat direction and
there is a term in the potential pushing the instanton size to zero.
This result, which will be important later, is in agreement with the
fact that there is no single instanton solution to Yang-Mills theory
on $T^{4}$ \cite{Braam:1988qk,vanBaal:1995eh}. There is a simple
argument that no such solution exists: if a $U(N)$ $k=1$ self-dual
solution existed on $T^{4}$ then the Nahm transformation would imply
the existence of a $k=N$ $U(1)$ solution on the dual torus! Since
it is obvious that no such solution can exist neither can the single
charge $U(N)$ configuration.

\section{The small instanton transition in type II brane world models}

So far we have described the dissolution of D3-branes onto D7-branes
in a manner which, apart from the additional contribution due to the
finiteness of the compact space, is conventional and in particular
is couched in ${\cal{N}}=2$ language. In order to make contact with phenomenological
{}``brane world'' models, we have to deal with two things. First
we need to go to the more usual ${\cal{N}}=1$ language of phenomenological
models, whilst showing that the instanton data sector of the theory
retains the essential underlying ${\cal{N}}=2$ structure. Our second problem
arises from the fact that the system is not stabilised and therefore
we have to allow the moduli fields to become full dynamical degrees
of freedom. In particular the couplings $g_{D_{3}}$ and $g_{D_{7}}$
are as we have stressed functions of the moduli and they will change
during the cosmological evolution%
\footnote{An alternative approach would of course be to try to stabilise the
moduli somehow. However calculability would inevitably be lost that
way. In the interests of learning a bit more about the phase transition
itself, we prefer to consider a system of moving moduli that represents
a {}``pre-stabilised'' cosmology.%
}. This means we have to include in the action all of the moduli dependence
arising from the K\"ahler potential of the ${\cal{N}}=1$ theory. 

To do this we continue using the type IIB D3/D7-brane picture, and
first recall some basic facts about the effective theories of the
${\cal{N}}=1$ models. (The descriptions of the associated processes in type
IIA and in G2 compactifications of M-theory, as described in the introduction,
can then be obtained by considering the mirror of our discussion and
its S-dual.) There are numerous string phenomenological models in
the literature which are specifically based upon D3 and D7-brane configurations
on orientifolds of Calabi-Yau spaces in type IIB 
\cite{blumreview,branepheno37}. We shall
not concern ourselves with any one particular such vacuum as we wish to keep
our considerations as general as possible. 
Instead we shall just require that the eventual four dimensional 
theory has ${\cal{N}}=1$ supersymmetry, and will achieve this by 
compactifying on a space which is an orientifold
of an appropriate orbifold, with a covering space that is a product of 
2-tori. We will assume that there is a stack of 2 D7-branes
wrapping an appropriate 4 cycle which for simplicity 
does not lie on or pass
through any fixed planes of the orientifold. We choose to label the
three $T^{2}$ of the orbifold such that these D7-branes wrap tori
1 and 2 and are transverse to torus 3. We shall also require that
there is a single D3-brane close to this stack of D7-branes lying
in the four dimensions of our space and time. Given this minimal set-up 
the following ${\cal{N}}=1$ structure describes the relevant moduli in
the four dimensional effective description of the system \cite{Ibanez:1998rf,Grana:2003ek,Lust:2004cx,Lust:2004fi,Jockers:2004yj,Jockers:2005zy}:

\begin{eqnarray}
K & = & -\ln(S+\bar{S)}-\sum_{i=1}^{3}\ln(T_{i}+\bar{T_{i})}+\sum_{i=1}^{3}\frac{|C_{3}^{i}|^{2}}{T_{i}+\bar{T_{i}}}\nonumber \\
 &  & +\sum_{i=1,j\neq i}^{2}\frac{|C_{i}^{7}|^{2}}{T_{j}+\bar{T}_{j}}+\frac{|C_{3}^{7}|^{2}}{S+\bar{S}}+\frac{1}{2}\frac{(|A|^{2}+|B|^{2})}{\sqrt{(T_{1}+\bar{T}_{1})(T_{2}+\bar{T}_{2})}}\end{eqnarray}

\begin{equation}
W=g_{3}(C_{1}^{3}C_{2}^{3}C_{3}^{3}+C_{3}^{3}AB)+g_{7}(C_{1}^{7}C_{2}^{7}C_{3}^{7}+AC^{7}B)\end{equation}

\begin{eqnarray}
h_{D3} & = & S\\
h_{D7\,\alpha\beta} & = & T_{3}\,\delta_{\alpha\beta}\end{eqnarray}
Here $g_{3}^{2}=2\pi$ and $g_{7}^{2}=\alpha^{'2}(2\pi)^{5}$ (not
to be confused with the effective couplings in the $D$-terms $g_{D_{3}}$
and $g_{D_{7}}$ which we shall determine shortly) and the $h$'s are
the gauge kinetic functions. In these expressions $S$ is the dilaton
superfield and $T_{i}$ are the superfields associated with the sizes
of the three 2-tori of the orbifold. The $C^i_{3}$ superfields are
associated with the position of the D3-brane within the compact space
and the $C^{7}_i$ moduli describe the D7-brane position moduli and the
moduli associated with various Wilson lines on the stack's world volume.
Finally the superfields $A$ and $B$ are associated with twisted
states of strings stretching between the D3 and D7-branes. These states
are massless when the D3/D7-branes are coincident and indeed the mass
is proportional to the stretching distance between D3 and D7-branes. 

Next we wish (with minimal loss of generality) to truncate the field
content to reduce the number of fields we have to consider. However,
we still must keep those necessary to describe the phenomena we want
to investigate. This truncation must of course be consistent in that
a solution to the truncated system should also be a solution to the
full one. A consideration of the above shows that the following truncations
work:

\begin{equation}
C_{3}^{1}=C_{3}^{2}=0,\,\,\,\,\, C_{1}^{7}=C_{2}^{7}=0,\,\,\,\,\, T_{1}=T_{2}\equiv T\,.\end{equation}

The first of these corresponds to not exciting the D3-brane's motion
parallel to the stack of D7-branes. The second set of constraints
corresponds to dropping certain Wilson line moduli associated with
the D7 stack. For brane stacks of different dimensionalities it may
be of interest to keep such moduli in future solutions, as keeping
more than one of these matrix valued brane coordinates could give 
information about the
onset of fuzziness due to cosmological evolution. The
final truncation just corresponds to using an obvious symmetry between
the $T$ superfields associated with the two $T^{4}$ wrapped by the
D7 to simplify our equations. Performing these truncations our system
becomes

\begin{equation}
K_{1}=-\ln(S+\bar{S)}-2\ln(T+\bar{T})-\ln(T_{3}+\bar{T_{3})}+\frac{|C_{3}|^{2}}{T_{3}+\bar{T}_{3}}+\frac{|C^{7}|^{2}}{S+\bar{S}}+\frac{1}{2}\frac{(|A|^{2}+|B|^{2})}{T+\bar{T}}\end{equation}

\begin{equation}
W_{1}=g_{3}(C_{3}AB)+g_{7}(AC^{7}B)\end{equation}

\noindent where we have dropped the indices on the remaining $C_{3}$
and $C^{7}$ fields as this now causes no ambiguity. It should be
noted that we are also suppressing indices describing how $A$,$B$,$C_{3}$
and $C^{7}$ transform under the D-brane gauge groups. The twisted
sector states transform under bi-fundamental representations of the
two unitary groups and $C_{3}$ and $C^{7}$ are of course in the
adjoint of the D3 and D7 gauge groups respectively. Thus the only
ambiguity that arises in our suppression of these indices above is
in the $AC^{7}B$ term. Here the indices in the fundamental of $U(2)_{D7}$
contract as $A_{\alpha}C_{\,\,\,\,\,\,\,\,\beta}^{7\,\alpha}B^{\beta}$
and from now on it will be useful to use the expansion $C_{\,\,\,\,\,\,\,\beta}^{7\,\alpha}=C^{7\,(a)}t_{\,\,\,\,\beta}^{(a)\,\,\,\alpha}$
where the $t$ are the generators of $U(2)$. More generally in the
$U(N)$ case the $t_{(a)}$ are the D7 stack gauge group generators.

We can now turn to the component action form of the scalar
sector of the full ${\cal{N}}=1$ supergravity theory; we find

\begin{eqnarray}
S & = & \int\sqrt{-g}\left[\left(\frac{1}{(S+\bar{S})^{2}}+\frac{2|C^{7\,(a)}|^{2}}{(S+\bar{S})^{3}}\right)\partial S\partial\bar{S}-\frac{1}{(S+\bar{S})^{2}}\left(\partial SC^{7\,(a)}\partial\bar{C}^{7\,(a)}+\partial\bar{S}\bar{C}^{7\,(a)}\partial C^{7\,(a)}\right)\right.\nonumber \\
 &  & +\left(\frac{1}{(T_{3}+\bar{T}_{3})^{2}}+\frac{2|C_{3}|^{2}}{(T_{3}+\bar{T}_{3})^{3}}\right)\partial T_{3}\partial\bar{T}_{3}-\frac{1}{(T_{3}+\bar{T}_{3})^{2}}\left(\partial T_{3}C_{3}\partial\bar{C}_{3}+\partial\bar{T}_{3}\bar{C}_{3}\partial C_{3}\right)\nonumber \\
 &  & +\left(\frac{2}{(T+\bar{T})^{2}}+\frac{(|A|^{2}+|B|^{2})}{(T+\bar{T})^{3}}\right)\partial T\partial\bar{T}-\frac{1}{2}\frac{1}{(T+\bar{T})^{2}}\left(\partial T(A\partial\bar{A}+\partial\bar{B}B)+\partial\bar{T}(\partial A\bar{A}+\bar{B}\partial B)\right)\nonumber \\
 &  & \left.+\frac{1}{T_{3}+\bar{T}_{3}}\partial C_{3}\partial\bar{C}_{3}+\frac{1}{S+\bar{S}}(\partial C^{7\,(a)}\partial\bar{C}^{7\,(a)})+\frac{1}{2}\frac{1}{T+\bar{T}}(\partial A\partial\bar{A}+\partial\bar{B}\partial B)+V_{F}+V_{D}\right]\end{eqnarray}
In the above $V_{F}$ and $V_{D}$ are the contributions to the potential
from ${\cal{N}}=1$ $F$-terms and $D$-terms respectively. In our discussion
we will only be working to fourth order in the matter fields (as well
as applying the slowly moving modulus approximation). In this 
approximation we find
the following expression for $V_{F}$ from the standard ${\cal{N}}=1$ structure:

\begin{eqnarray}
V_{F} & = & e^{K}\, K^{-1\,\bar{i}j}\bar{F}_{\bar{i}}F_{j}\nonumber \\
 & = & \frac{1}{(S+\bar{S})(T+\bar{T})^{2}(T_{3}+\bar{T}_{3})}\left[g_{3}^{2}|AB|^{2}(T_{3}+\bar{T}_{3})+g_{7}^{2}\sum_{(a)}|At^{(a)}B|^{2}(S+\bar{S})\right.\nonumber \\
 &  & +\left\{ \left(g_{3}\bar{C}_{3}\bar{B}+g_{7}\bar{C}^{7\,(a)}\bar{B}t^{(a)}\right)\left(g_{3}C_{3}B+g_{7}C^{7\,(b)}t^{(b)}B\right)\right.\nonumber \\
 &  & +\left.\left.\left(g_{3}C_{3}A+g_{7}C^{7\,(a)}At^{(a)}\right)\left(g_{3}\bar{C}_{3}\bar{A}+g_{7}\bar{C}^{7\,(b)}t^{(b)}\bar{A}\right)\right\} 2(T+\bar{T})\right]\, . \end{eqnarray}
For $V_{D}$ we have\begin{equation}
V_{D}=\frac{1}{2}Re(h_{D3})^{-1}D^{(D3)}D^{(D3)}+\frac{1}{2}Re(h_{D7^{(a)}})^{-1}D^{(D7^{(a)})}D^{(D7^{(a)})}\end{equation}
 where, using the usual formulae for the $D$-terms, we have 

\[
D^{(D3)}=\frac{g_{3}}{2}\frac{(|B|^{2}-|A|^{2})}{(T+\bar{T})}\]

\begin{equation}
D^{(D7^{(a)})}=\frac{g_{7}}{2}\frac{(At^{(a)}\bar{A}-\bar{B}t^{(a)}B)}{T+\bar{T}}+\frac{g_{7}f_{abc}\bar{C}^{7\,(b)}C^{7\,(c)}}{S+\bar{S}}\end{equation}
where $f_{abc}$ are the $U(N)$ structure constants.

\subsection*{\textmd{\emph{Connection with the}} \textmd{${\cal{N}}=2$} \textmd{\emph{structure }}}

Before proceeding to make further truncations it will now be helpful
for us to show explicitly how the ${\cal{N}}=2$ structure and potential described
in the previous section is embedded in this result. In order to do 
this we must first identify
the ${\cal{N}}=2$ hypermultiplets associated with the instanton data. These
fields must be composed out of a non-holomorphic combination of fields
from the ${\cal{N}}=1$ viewpoint as the $SU(2)$ automorphism index which
they carry is associated with the different possible choices of complex
structure associated with the ${\cal{N}}=2$ theory; clearly if we are writing
things in terms of the ${\cal{N}}=1$ language then we have chosen one particular
such complex structure. In fact, given that $A$ and $\bar{B}$ have
the same charge under the $U(2)$ associated with the D7 stack%
\footnote{The charges are as follows 

\begin{center}\begin{tabular}{|c|c|c|c|}
\hline 
$field$&
$Q_{3}$&
$Q_{7}$&
$\sigma_{3}(SU(2)_{\textnormal{auto}})$\tabularnewline
\hline
\hline 
$A_{1}$&
1&
-1&
+$\frac{1}{2}$\tabularnewline
\hline 
$A_{2}$&
1&
-1&
$-\frac{1}{2}$\tabularnewline
\hline 
$B^{1}$&
-1&
1&
-$\frac{1}{2}$\tabularnewline
\hline 
$B^{2}$&
-1&
1&
+$\frac{1}{2}$\tabularnewline
\hline
\end{tabular}\end{center}%
}, we are led to suggest the identification $\chi_{1\,\alpha}=A_{\alpha}$
and $\chi_{2\,\alpha}=\bar{B}_{\alpha}.$

To show this identification is correct we will now recover the potential
given in the previous section from that given above in the theory
derived from the ${\cal{N}}=1$ structure. The D-terms in the ${\cal{N}}=2$ potential
come in the ${\cal{N}}=1$ language from both $D$ and $F$ terms where, to
obtain parity with the previous section, we set the metric moduli
and dilaton to be constant. The contribution to the potential given
above coming from the sum of the $D^{2}$ term of the D3-brane and 
the $|F_{C_{3}}|^{2}$ term is \begin{equation}
V_{D3}=\frac{g_{3}^{2}(4|AB|^{2}+(|A|^{2}-|B|^{2})^{2})}{4(S+\bar{S})(T+\bar{T})^{2}}.\end{equation}
To show that this is compatible with our identification of the ${\cal{N}}=2$
structure, we use, \begin{eqnarray}
\sum_{A}^{3}(\chi_{i\gamma}^{\dagger}\sigma_{ij}^{A}\chi_{j\gamma})^{2} & = & 4|\bar{\chi}_{1}\chi_{2}|^{2}+(\bar{\chi}_{1}\chi_{1}-\bar{\chi}_{2}\chi_{2})^{2}\nonumber \\
 & = & 4|AB|+(|A|^{2}-|B|^{2})^{2}\end{eqnarray}
where we have suppressed contracted gauge group indices as elsewhere
in this section. This contribution to the potential can then be written
as\begin{equation}
V_{D3}=\frac{g_{3}^{2}}{4(S+\bar{S})(T+\bar{T})^{2}}\,\sum_{A}^{3}(\chi_{i\gamma}^{\dagger}\sigma_{ij}^{A}\chi_{j\gamma})^{2}.\end{equation}
If we identify $g_{D_{3}}=g_{3}/\sqrt{(S+\bar{S})(T+\bar{T})^{2}}$,
we then see that this term reproduces the contribution to the potential
coming from the ${\cal{N}}=2$ $D$-term associated with the D3-brane, 
as seen in the previous section. Note
that the correct identification relies on specific couplings in the
${\cal{N}}=1$ superpotential and also on the correct relation between the
K\"ahler potential and gauge kinetic function. 

Similarly the contribution to our potential from the $D^{2}$ terms
from the D7-branes, plus the $|F_{C_{7}}|^{2}$ term is \begin{equation}
V_{D7}=\frac{g_{7}^{2}(\sum_{a=1}^{4}4|At^{(a)}B|^{2}+\sum_{a=1}^{4}(At^{(a)}\bar{A}-\bar{B}t^{(a)}B)^{2})}{4(T+T)^{2}(T_{3}+\bar{T}_{3})}.\end{equation}
Comparing this with \begin{eqnarray}
\sum_{A}^{3}(\chi_{i\alpha}^{\dagger}t_{\,\,\,\,\,\beta}^{(a)\,\alpha}\sigma_{ij}^{A}\chi_{j}^{\beta})^{2} & = & 4|\bar{\chi}_{1}t^{(a)}\chi_{2}|^{2}+(\bar{\chi}_{1}t^{(a)}\chi_{1}-\bar{\chi}_{2}t^{(a)}\chi_{2})^{2}\nonumber \\
 & = & 4|At^{(a)}B|^{2}+(At^{(a)}\bar{A}-\bar{B}t^{(a)}B)^{2}\end{eqnarray}
we find the expression \begin{equation}
V_{D7}=\frac{g_{7}^{2}}{4(T+\bar{T})^{2}(T_{3}+\bar{T}_{3})}\,\sum_{a=1}^{4}\sum_{A}^{3}(\chi_{i\alpha}^{\dagger}t_{\,\,\,\,\,\beta}^{(a)\,\alpha}\sigma_{ij}^{A}\chi_{j}^{\beta})^{2}.\end{equation}
Thus defining $g_{D7}=g_{7}/\sqrt{(T+\bar{T})^{2}(T_{3}+\bar{T}_{3})}$
we find that these terms reproduce the contribution to the potential
of the previous section from the ${\cal{N}}=2$ D7 D-terms.

\subsection*{\textmd{\emph{The action for cosmology}}}

Having confirmed the embedding of the ${\cal{N}}=2$ structure of the previous
section within the ${\cal{N}}=1$ language, it is now straightforward to identify
the ${\cal{N}}=1$ fields that correspond to the collective coordinates of
an instanton: we simply use equation (\ref{inst}) and the relationship
we have established between the two descriptions. We find that if
the twisted sector fields take the form, \begin{eqnarray}
A_{\alpha} & = & \rho(-i\theta_{3}+\theta_{4},\,-i\theta_{1}-\theta_{2})\label{adhmn1}\\
B^{\alpha} & = & \rho(i\theta_{1}+\theta_{2},\,-i\theta_{3}+\theta_{4})\label{adhmn2}\end{eqnarray}
 where $\sum_{n=1}^{4}\theta_{n}^{2}=1$ then they describe the collective
coordinates of an instanton of size $\rho$ and with $SU(2)$ orientation
moduli given by the $\theta$'s.

The next step is to identify a consistent truncation of our theory
that includes $A$ and $B$ in the form given in equations (\ref{adhmn1})
and (\ref{adhmn2}), so that we can interpret some of our fields as
describing an instanton on the D7 stack in the compactified dimensions
which it wraps. Inspection of the component form of the scalar sector
of the theory shows that the following truncations are consistent
when taken together. Firstly we set $C^{7(a)}=0$ unless $a$ labels
the identity matrix generator. In other words we just keep the D7
modulus corresponding to the overall position of our stack of D-branes
and discard moduli which describe such phenomena as the stack separating.
We will also truncate numerous axions such as the imaginary parts
of the $S$, $T$ and $T_{3}$ superfields. We write $C^{7\,(1)}=R(\cos(\psi)+i\sin(\psi))$
and $C_{3}=r(\cos(\chi)+i\sin(\chi))$ and then truncate $\psi$ and
$\chi$. This last truncation corresponds to having both branes simply
moving along the radial direction from some arbitrarily chosen origin
in the two dimensional subspace of the orientifold transverse to the
D7-brane stack. Finally we make what we shall refer to as the ADHM
truncation which simply corresponds to setting $A$ and $B$ to the
values given above.

After these simplifications we are left with the following consistent
truncation of the original theory:

\begin{eqnarray}
S & = & \int\sqrt{-g}\left[\mathcal{R}+\frac{1}{4}(\partial\phi)^{2}+\frac{1}{4}(\partial\beta_{3})^{2}+\frac{1}{2}(\partial\beta)^{2}+\frac{1}{2}e^{-\phi}(\partial R)^{2}+\frac{1}{2}e^{-\beta_{3}}(\partial r)^{2}\right.\nonumber \\
 &  & \left.+\frac{1}{2}e^{-\beta}(\partial\rho)^{2}+\frac{3}{32}g_{7}^{2}\rho^{4}e^{-2\beta-\beta_{3}}+(g_{3}r+\frac{g_{7}}{2}R)^{2}\frac{1}{2}\rho^{2}e^{-\phi-\beta-\beta_{3}}\right]\label{eq:action}\end{eqnarray}
In the above we have reinstated the four dimensional Einstein-Hilbert
term and we have also used the following definitions:

\begin{eqnarray}
Re\{ S\} & = & e^{\phi}+\frac{1}{2}R^{2}\nonumber \\
Re\{ T\} & = & e^{\beta}+\frac{1}{2}\rho^{2}\nonumber \\
Re\{ T_{3}\} & = & e^{\beta_{3}}+\frac{1}{2}r^{2}.\end{eqnarray}
To summarise the fields appearing in the action: $\phi$ is interpreted
as the four dimensional dilaton; the fields $\beta$ and $\beta_{3}$
describe the size moduli associated with the spaces wrapped by and
transverse to the D7-brane stack respectively; 
the fields $R$ and $r$ describe radial motion of 
the D7 stack and D3-brane respectively relative
to some origin in the third torus of the orbifold;
finally $\rho$ describes the size of the instanton which results
when the D3-brane dissolves upon the stack of D7-branes%
\footnote{This physical interpretation of the fields in the action conflicts
somewhat with the results presented in \cite{Lust:2004cx,Lust:2004fi}
but is in agreement with the results of \cite{Jockers:2004yj}. We
have chosen this interpretation as it is the one which, on writing
out the complete component action instead of the truncation presented
above, preserves one of the necessary symmetries: the physics of the
system should not depend on the arbitrary choice of the origin with
respect to which the overall brane position moduli are measured. In
other words there should be a shift symmetry of the action associated
with the brane position coordinates, and for this it is important
to take into account the possibility that the metric moduli and dilaton
superfield definitions are modified when the brane position moduli
are included in the description of the system.%
}. Note that the potential found in this action has the properties
we would expect from the above discussion. In particular when $(g_{3}r+\frac{g_{7}}{2}R)=0$
so that we are in the Higgs branch the potential is proportional
to $g_{7}^{2}$ indicating that it comes from $F$ and $D$-terms associated
with the D7-brane and is proportional to the size of the instanton
to the fourth power, $\rho^{4}$. Thus as described in the previous
section we find that the flat direction corresponding to the size
of an instanton in the D3/D7-brane system in an infinite volume is
no longer flat after compactification. 

One question that could be asked about this {}``ADHM-like'' truncation
we have chosen is whether it is actually preferred in some sense.
To answer this we can examine the structure of the potential before
this truncation, by replacing $|A|^{2}=x$ , $|B|^{2}=y$ and $|A.B|=z$
and treating $x,y,z$ as independent. Using the standard $U(2)$ identity
$\frac{1}{2}\delta_{\beta}^{\alpha}\delta_{\delta}^{\gamma}=t_{\,\delta}^{(a)\alpha}t_{\,\beta}^{(a)\gamma}$
the total potential before the trunction is \begin{equation}
4V=(4g_{D3}^{2}-g_{D7}^{2})\, z^{2}+(g_{D3}^{2}+\frac{1}{2}g_{D7}^{2})\,(x^{2}+y^{2})+2(g_{D7}^{2}-g_{D3}^{2})\, x\, y\,.\end{equation}
The potential is unbounded in the $z$ direction if $g_{D7}>2g_{D3}$,
however that would require small $T_{3}$ so that the approximation
of D7-branes wrapping large dimensions would be invalid and one would
have to go to a $T$-dual description. Neglecting this possibility
and assuming that $g_{D7}\ll2g_{D3}$, we get an extremum at $x=y=z=0.$
In the $z$ direction this is independently a minimum. Calling $\lambda=g_{D7}^{2}/g_{D3}^{2}\ll1$
we see that the fields are driven to $z=0$ and $x=y$ by a gradient
of order $g_{D3}^{2}$, whereas they are driven to $x=y=0$ with a
gradient of order $\lambda g_{D3}^{2}$. Comparing with the form of
the ADHM truncation we then see that for small $\lambda$ we can assume
that we quickly roll to an ADHM-like solution. (For larger values
of $g_{D7}$ one would not expect the ADHM solution to be particularly
preferred however, and then it would be more complete to treat $x,y,z$
as independent fields).

\section{The dynamics of brane dissolution}

The action presented in eq.(\ref{eq:action}) describes
a system which can yield simple cosmological solutions. These can be
found analytically in some regimes but certainly numerically in
general; we now examine their behaviour. The equations of motion 
when an FRW ansatz is made for the $d=4$ metric 
are relatively straightforward to obtain and are presented in the Appendix. 

Firstly we consider solutions where the twisted sector fields are
initially set to zero. Making the usual FRW ansatz for the four dimensional
metric with flat spatial sections, $ds^{2}=-dt^{2}+e^{2\alpha}d\vec{x}^{2}$
, we find the following complete class of solutions:\begin{eqnarray}
\alpha & = & \frac{1}{3}\ln(t-t_{0})+\alpha_{0}\nonumber \\
\beta & = & p_{\beta}\ln(t-t_{0})+\beta_{0}\nonumber \\
\phi & = & p_{\phi}\ln\left|\frac{t-t_{0}}{t_{1}}\right|-2\ln\left[1+\left|\frac{t-t_{0}}{t_{1}}\right|^{p_{\phi}}\right]+\phi_{0}\nonumber \\
\beta_{3} & = & p_{\beta_{3}}\ln\left|\frac{t-t_{0}}{t_{2}}\right|-2\ln\left[1+\left|\frac{t-t_{0}}{t_{2}}\right|^{p_{\beta_{3}}}\right]+\beta_{30}\nonumber \\
R & = & (R_{i}-R_{f})\left(1+\left|\frac{t-t_{0}}{t_{1}}\right|^{p_{\phi}}\right)^{-1}+R_{f}\nonumber \\
r & = & (r_{i}-r_{f})\left(1+\left|\frac{t-t_{0}}{t_{2}}\right|^{p_{\beta_{3}}}\right)^{-1}+r_{f}\,.\end{eqnarray}
In these solutions $p_{\phi},p_{\beta},p_{\beta_{3}},t_{0},t_{1},t_{2},\alpha_{0},\beta_{0},\beta_{30},\phi_{0},R_{i},R_{f},r_{i},r_{f}$
are constants of integration. They are subject to the following constraints:

\begin{eqnarray}
\phi_{0} & = & -\ln\left(\frac{2}{(R_{i}-R_{f})^{2}}\right)\nonumber \\
\beta_{3} & = & -\ln\left(\frac{2}{(r_{i}-r_{f})^{2}}\right)\nonumber \\
\frac{2}{3} & = & \frac{1}{4}p_{\phi}^{2}+\frac{1}{4}p_{\beta_{3}}^{2}+\frac{1}{2}p_{\beta}^{2}\,.\end{eqnarray}
The value of the field $\rho$ in the above vanishes for all time.
These solutions take a form which is familiar to string cosmologists.
The solutions split up into two types, positive and negative time
branches, depending on what ranges one allows for $t$ consistent
with keeping the logarithms in the solutions well defined. Asymptotically,
both at early and late times, the D3 and D7-branes are stationary
and the remaining moduli are evolving in a standard rolling radius
solution \cite{Mueller:1989in,Brandle:2000qp}. Then at some intermediate
time, determined by the integration constant $t_{1}$, the D7 stack
suddenly and monotonically moves from its initial to its final position.
This behaviour should be compared with that of moving M5 branes in
heterotic M-theory, evolution of gauge fivebrane sizes and other bundle
moduli, and evolution of form field expectation values in various settings
\cite{Copeland:1994vi,Lidsey:1999mc,Copeland:2001zp,Gray:2003vk,Gray:2004rw}.
Similarly the D3-brane moves at time $t-t_{0}=t_{2}$. The presence
of two times, $t_{1}$ and $t_{2}$, at which non-trivial evolution
takes place is in contrast to some previous solutions that have been
written down where all of the motion takes place at once.
This difference is due to the fact that the two kinds of D-brane have
different couplings to the remaining moduli fields and in general
one should expect one such integration constant for each different
coupling in the system. 

\begin{figure}
\begin{center}\begin{tabular}{cc}
\includegraphics[%
  scale=0.5]{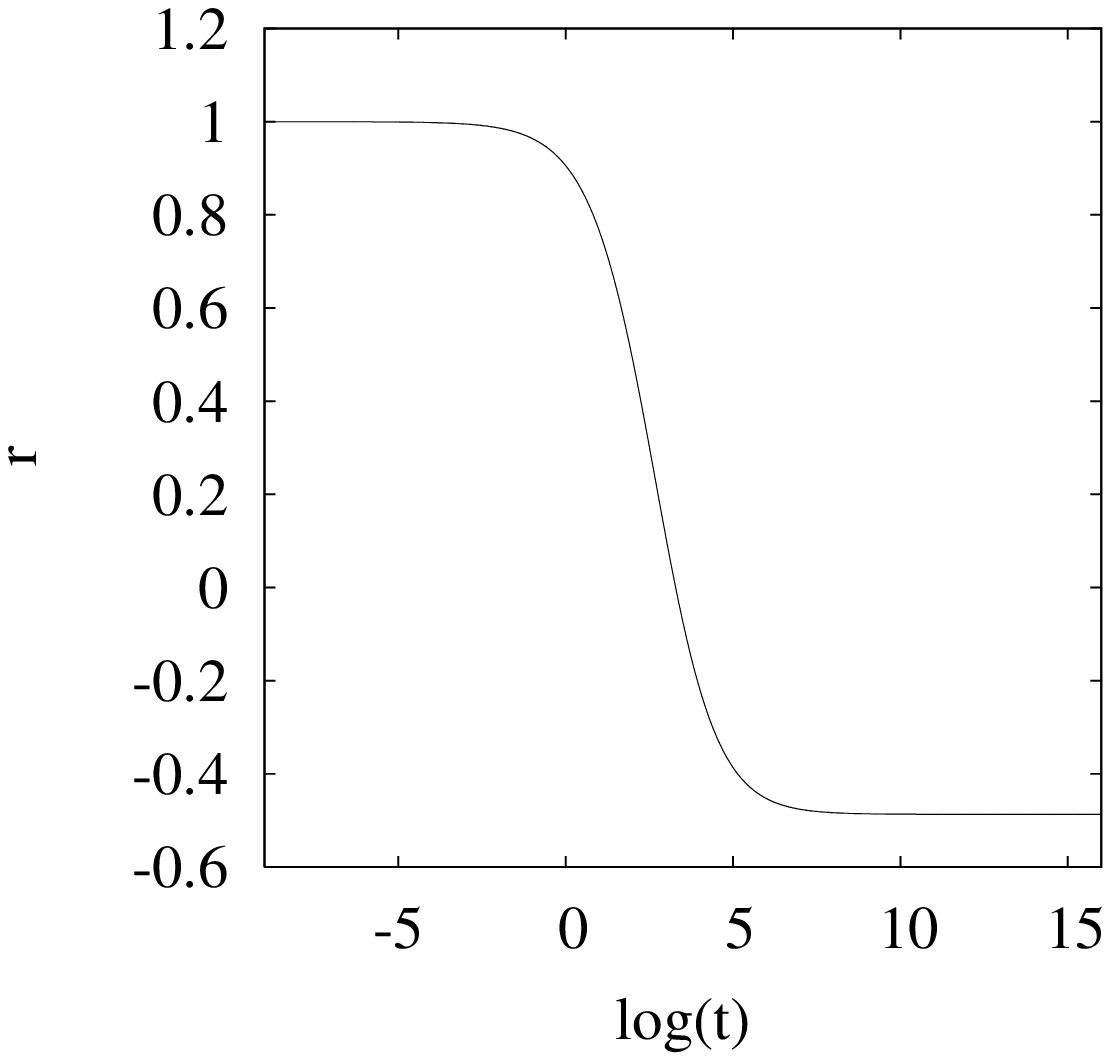}&
\includegraphics[%
  scale=0.5]{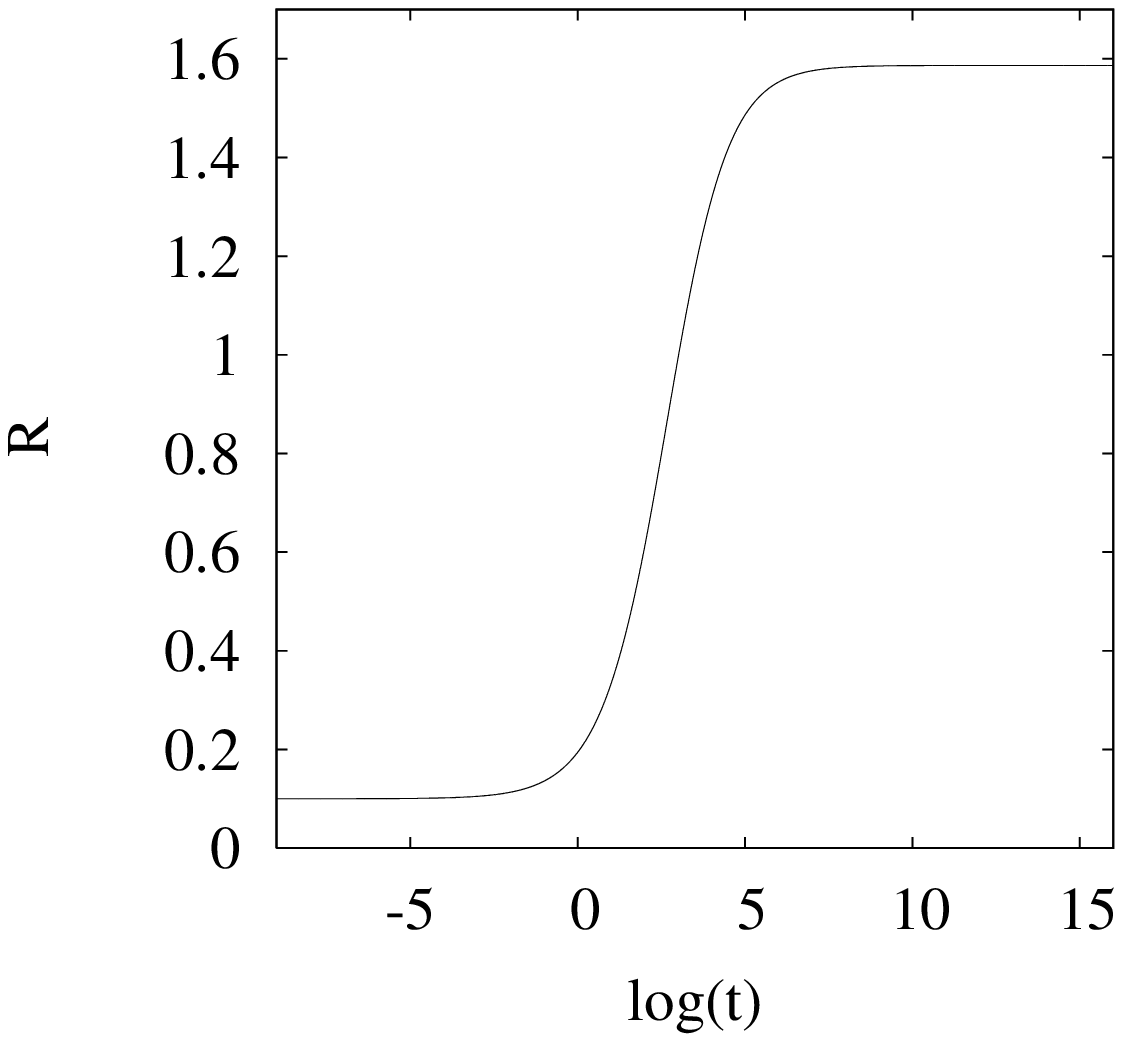}\tabularnewline
\includegraphics[%
  scale=0.5]{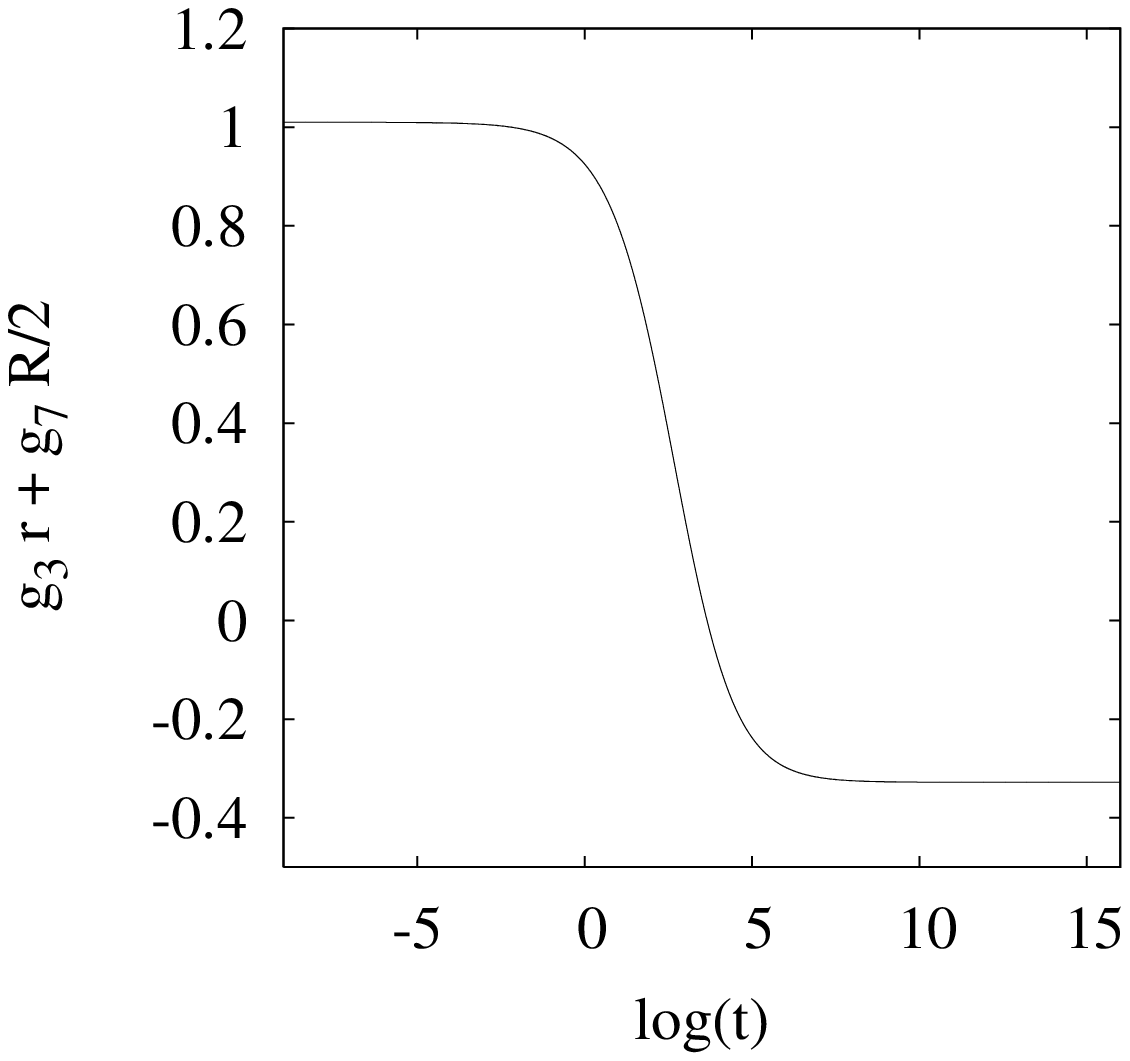}&
\includegraphics[%
  scale=0.5]{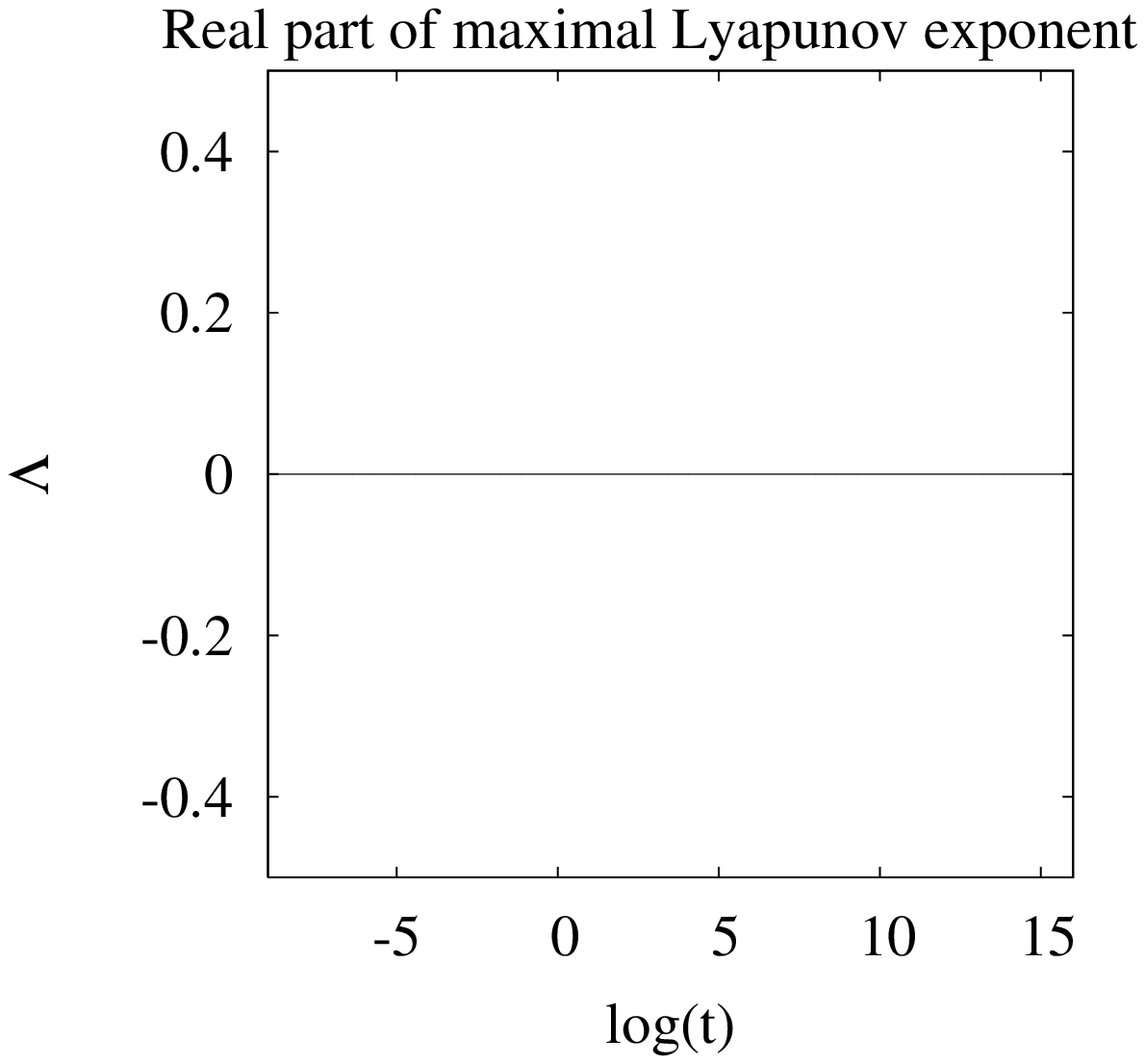}\tabularnewline
\end{tabular}\end{center}

\caption{\label{cap:no-oscillation}A solution in which
the D3-brane passes straight through the D7 stack. The $x$-axis
corresponds to time plotted on a log scale. The 
combination $g_3 r+\frac{g_7}{2}R$ corresponds to
the D3/D7-brane separation.}
\end{figure}

The plots in figure~\ref{cap:no-oscillation} show some of the features
of these solutions. In particular we have chosen our integration constants
such that the D3 and D7-branes pass through each other. These plots
were generated using a numerical solution to the equations of motion
in order to check the numerical algorithm against an available analytical
result. It should be noted that no small instanton transition occurs.
The reason for this is obvious on inspection of the action in eq.(\ref{eq:action}).
One can see that it is a consistent truncation to set the twisted
sector fields $\rho$ to zero; in so doing these fields are completely
removed from the action and if they are zero initially then they remain
so for all time.

From this analysis one might conclude that, since the twisted sector
fields are massive when the D3 and D7-branes are separated, small
instanton transitions occur extremely rarely. However this is not
the case because, as we shall see, even relatively small excitations
of the twisted sector - consistent with their massive nature for separated
D-branes - can result in small instanton transitions when the branes
come together. 

\begin{figure}
\begin{center}\begin{tabular}{c}
\begin{tabular}{cc}
\includegraphics[%
  scale=0.5]{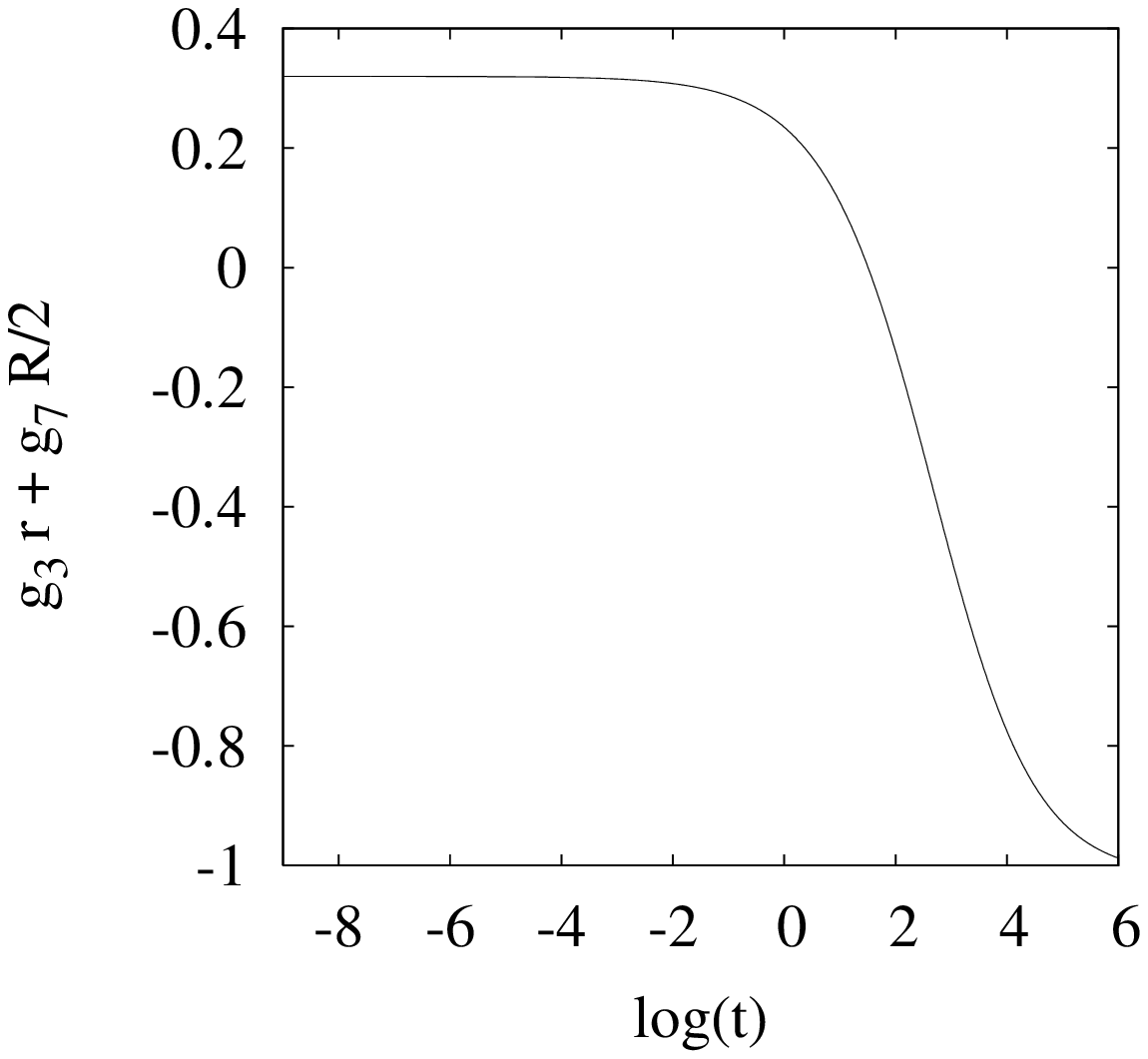}&
\includegraphics[%
  scale=0.5]{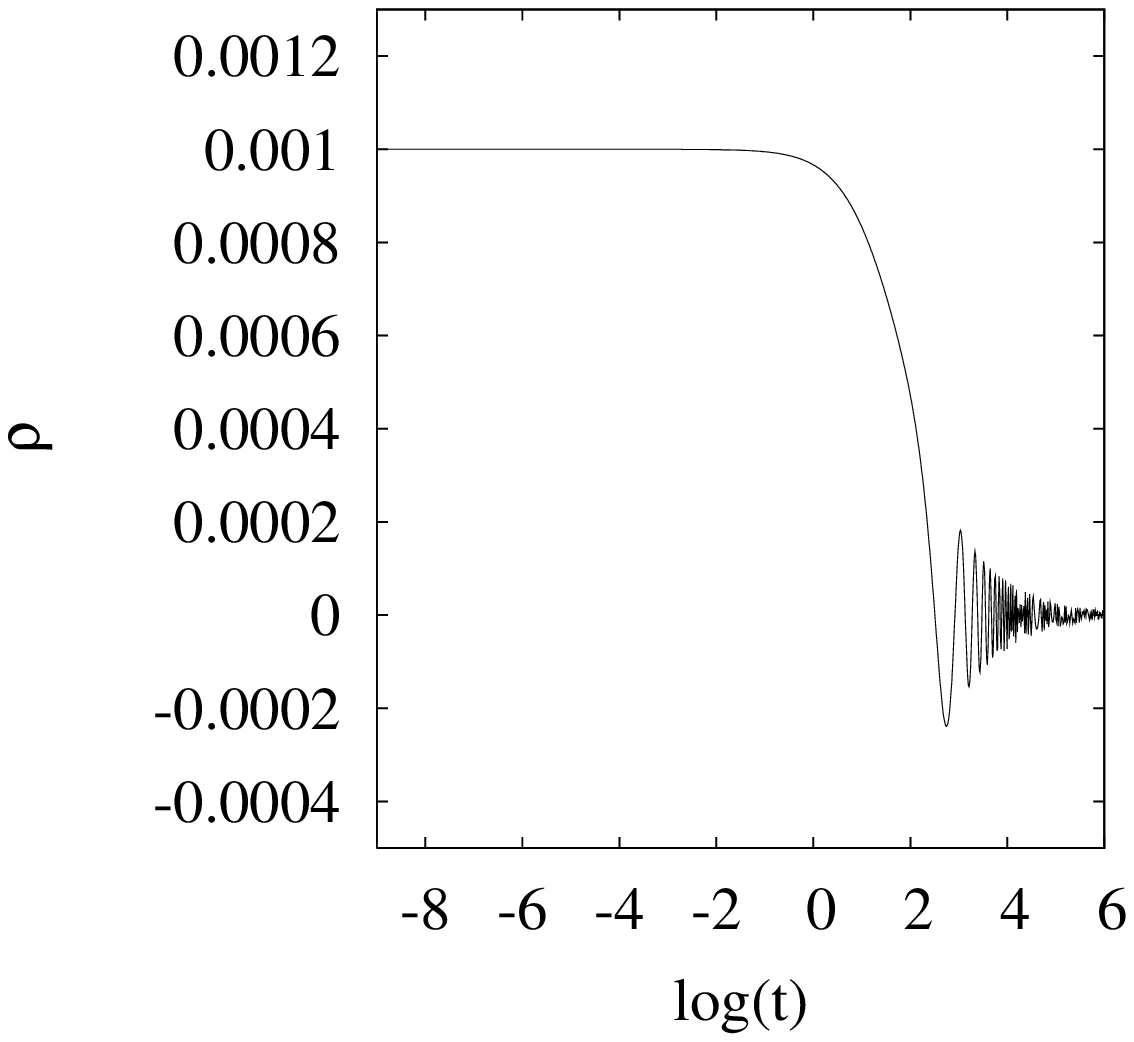}\tabularnewline
\end{tabular}\tabularnewline
\includegraphics[%
  scale=0.5]{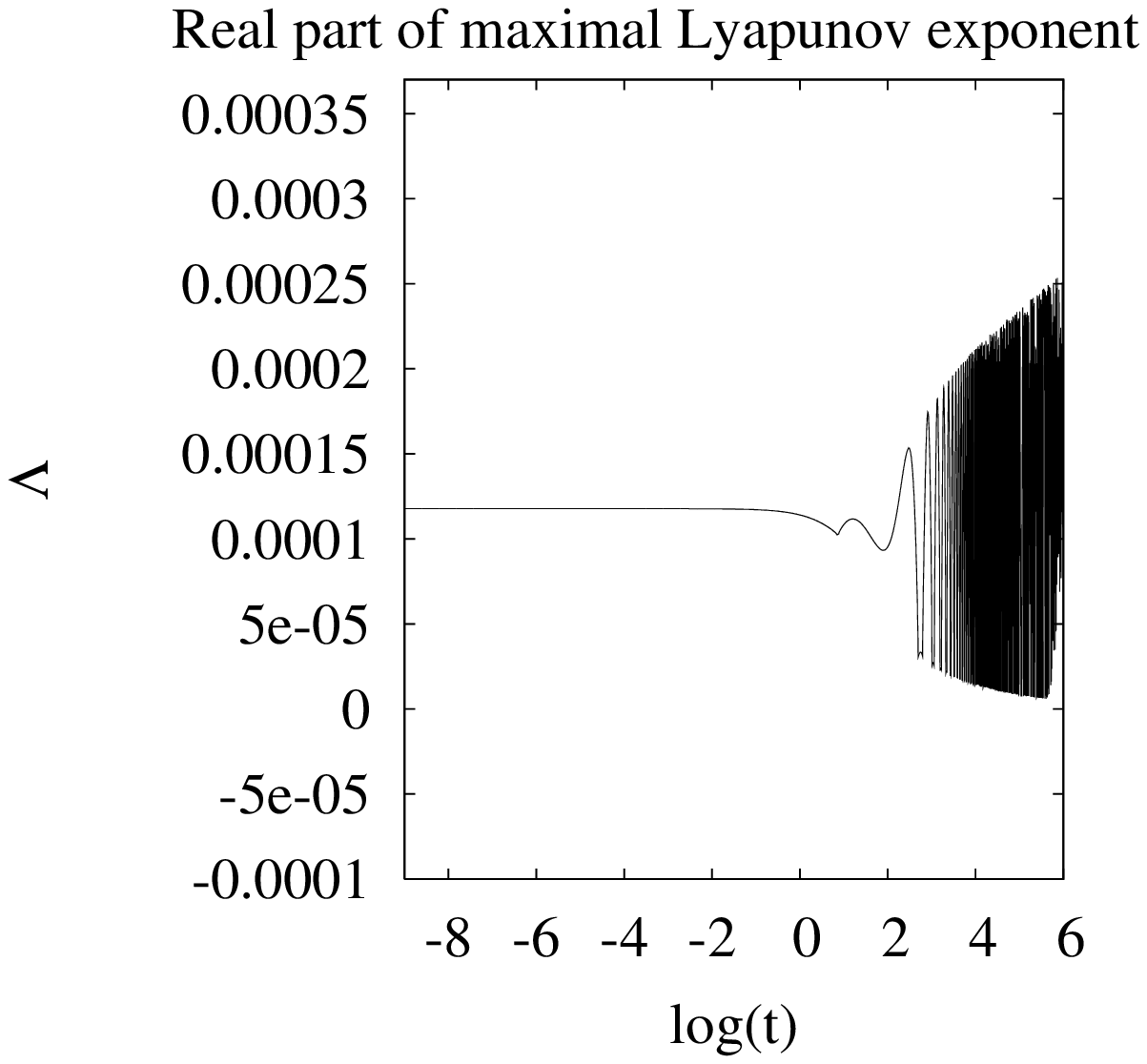}\tabularnewline
\end{tabular}\end{center}

\caption{\label{cap:caption-number-two}A slightly excited D3-brane still passes
straight through the D7-brane. As before time is plotted on a log
scale.}
\end{figure}

Let us now consider adding excitations of the twisted sector to our
initial conditions in an effort to induce a small instanton transition.
The resulting dynamics can be obtained numerically. As figure~\ref{cap:caption-number-two}
shows, for a very small initial excitation the twisted sector field
simply oscillates and decays with time; the branes still pass
straight through each other and no brane dissolution occurs.

This behaviour can be understood from the form of the potential for
the system in the $\rho$, $g_{3}r+\frac{g_{7}}{2}R$ plane. The minima
of the potential forms a cross shape, one branch corresponding to
$\rho$ being zero (the Coulomb phase with arbitrary separation of
the D3 and D7-branes), and the other branch corresponding to the D3
brane lying on top of the D7-branes and a non-zero size of the dissolved
D3-brane (the Higgs branch). As we have seen, at large compactification
volumes this branch is not quite flat due to the $\rho^{4}$ piece
in the potential. Our initial conditions here have started the system
far along the Coulomb branch with $\rho$ slightly excited up the
potential. The twisted sector field thus undergoes oscillations, which
are Hubble damped, as the D-brane separation heads towards zero. As
the system traverses the hub of the cross, if the excitations of $\rho$
are not large enough, it continues down the other side of the \emph{same
branch} it started on.

\begin{figure}
\begin{center}\includegraphics[%
  width=3.5in,
  height=2.5in]{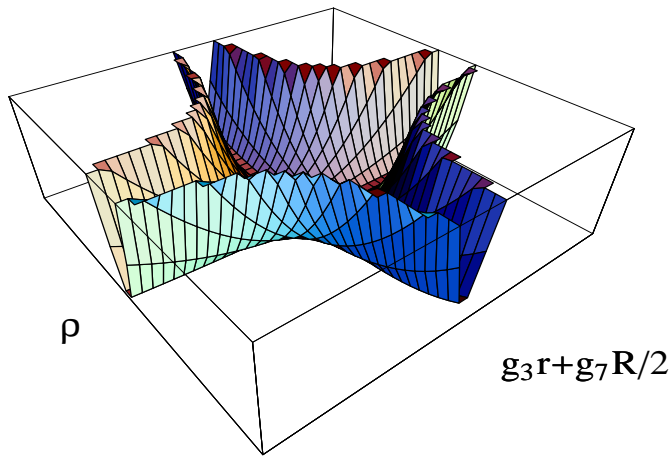}\end{center}

\caption{The potential}
\end{figure}

\begin{figure}
\begin{center}\begin{tabular}{c}
\begin{tabular}{cc}
\includegraphics[%
  scale=0.5]{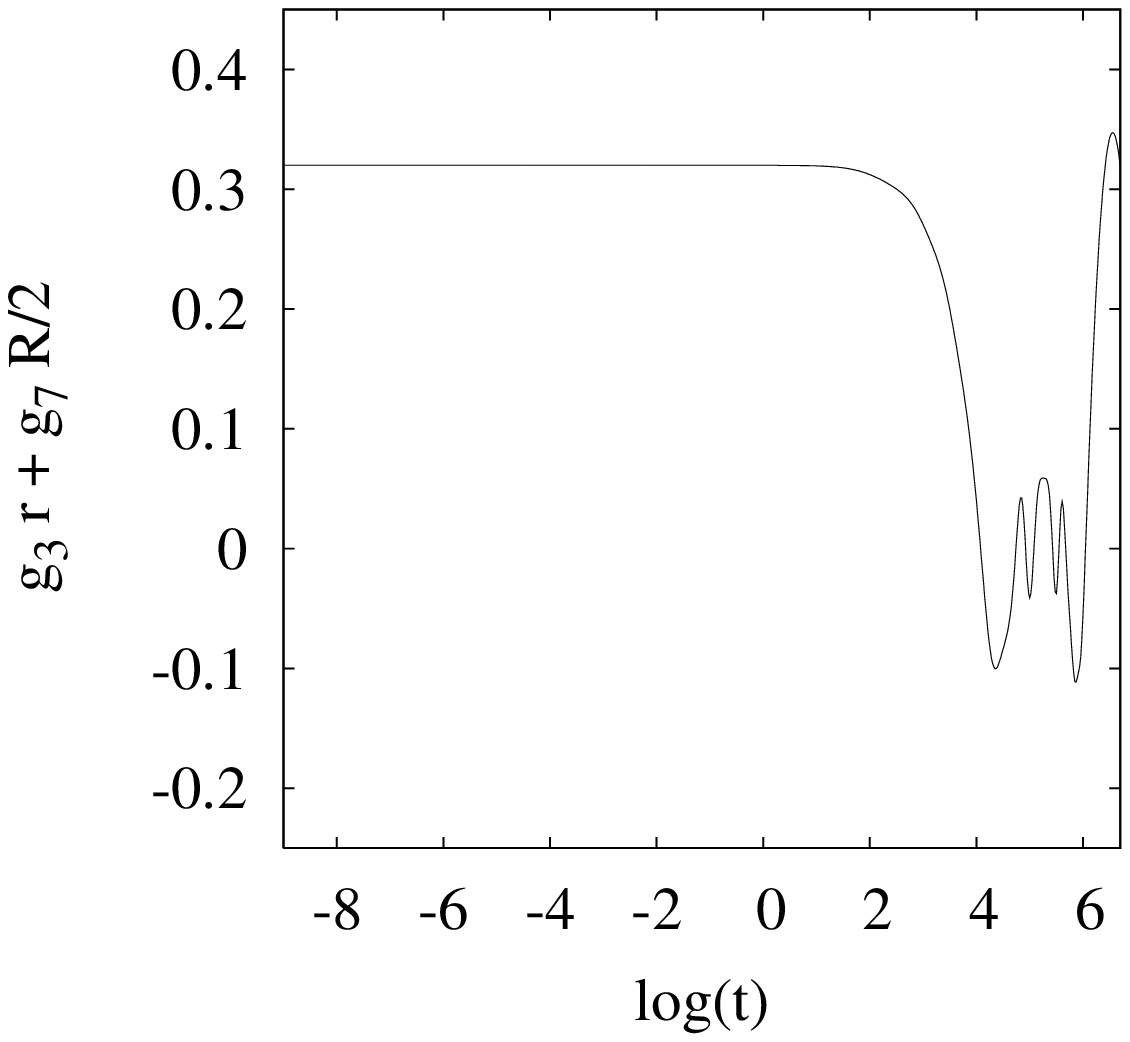}&
\includegraphics[%
  scale=0.5]{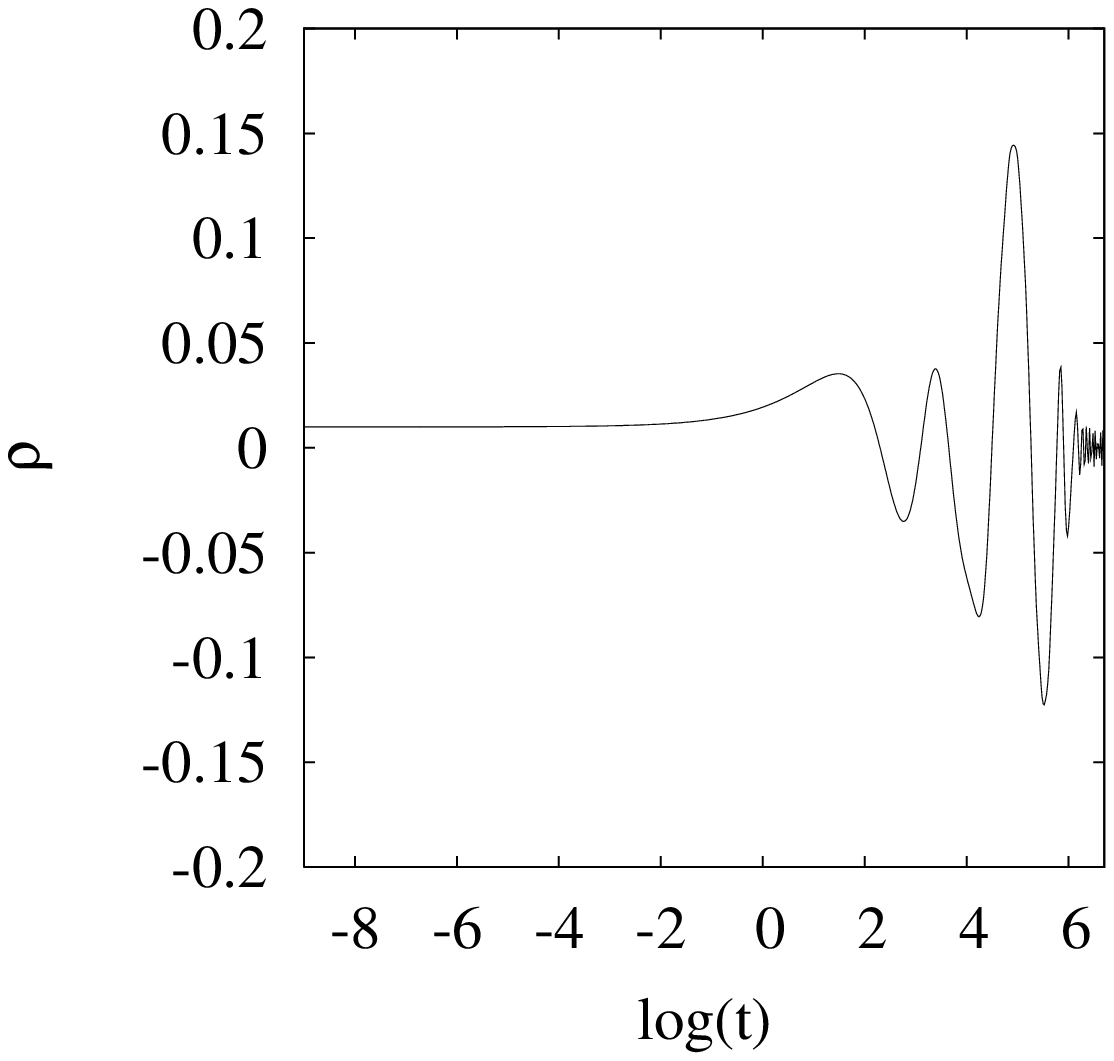}\tabularnewline
\end{tabular}\tabularnewline
\includegraphics[%
  scale=0.5]{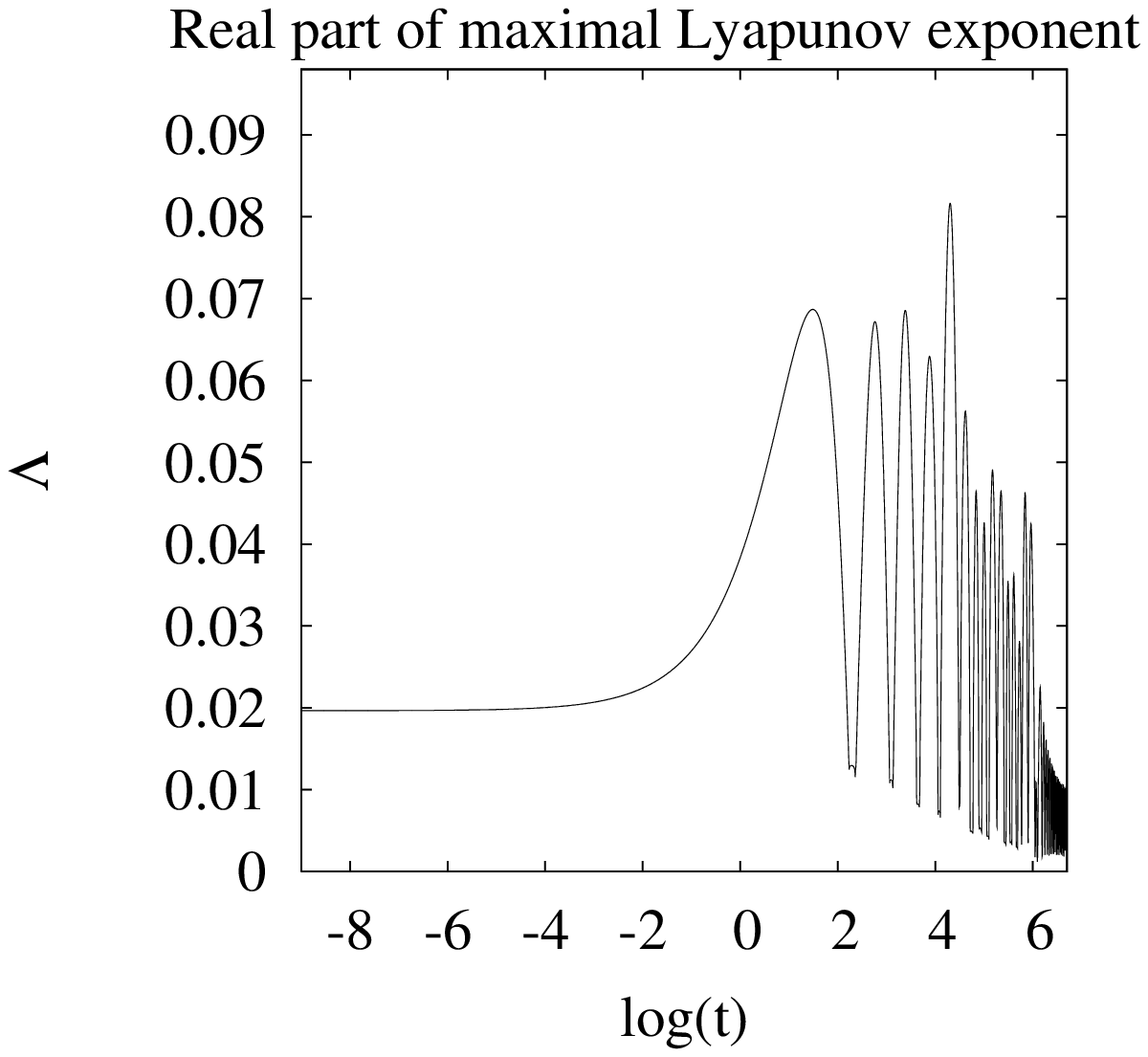}\tabularnewline
\end{tabular}\end{center}

\caption{A sequence of two small instanton transitions: a D3-brane is captured
but later emitted. As before time is plotted on a log scale.\label{onagainoffagain}}
\end{figure}

Given this understanding of the system's motion, exciting the twisted
sector further should result in the oscillations
being large enough to effect a branch change when the system reaches
the hub of the cross shaped potential %
\footnote{It should be noted that when we use the phrase {}``branch change''
here we are not referring to a change from cosmological contraction
to expansion as has been considered in some parts of the cosmological
literature in similar contexts \cite{cosmobranch}. Our solutions never exhibit
such a change. However the extra light states which appear here are
different in nature from those associated with a small instanton transition
in heterotic M-theory (as has already been mentioned). We thus can
not comment on whether such a {}``bouncing'' cosmology may be possible
in that case.%
}. This corresponds to the D3-brane dissolving onto the D7 stack upon
collision and does indeed occur as shown by the numerical results
in figure~\ref{onagainoffagain}. In this example the D3-brane is later
re-emitted. The reason for this and the details of the late time behaviour
shown in these plots will be discussed in detail later.
For now we simply note that a small instanton transition has occurred
(although the instanton has not monotonically smoothed out to some
final size).%

Next we can ask what happens when we start with an initially dissolved
D3-brane (i.e. with non-zero $\rho$) and allow it to evolve. If the
D-branes are all initially coincident and stationary, then they will
remain so for all time; again the relevant fields can be consistently
truncated from the system by giving them these initial conditions.
One then has a scalar field evolving in a quartic potential and one
may ask if the system inflates. Alas, the answer is not to any appreciable
degree. This is because the quartic potential has prefactors that
go as inverse powers of the sizes of the compact spaces. The potential
thus has two effects: it drives these moduli to large values and it
drives the twisted states to small vevs. The former effect occurs
so quickly due to the exponential nature of the relevant field dependence
that very little inflation can occur because the potential for the
twisted fields flattens out very quickly. This problem could perhaps
be solved by employing some stabilisation mechanism to fix the relevant
metric moduli fields. The potential would then still have various
problems however as large initial field values would be required to
obtain the requisite number of e-foldings, although such objections
may be resolvable in various ways. These are problems which 
are familiar in traditional chaotic inflation \cite{Linde:1983gd}. 
\begin{figure}
\begin{center}\begin{tabular}{c}
\begin{tabular}{cc}
\includegraphics[%
  scale=0.5]{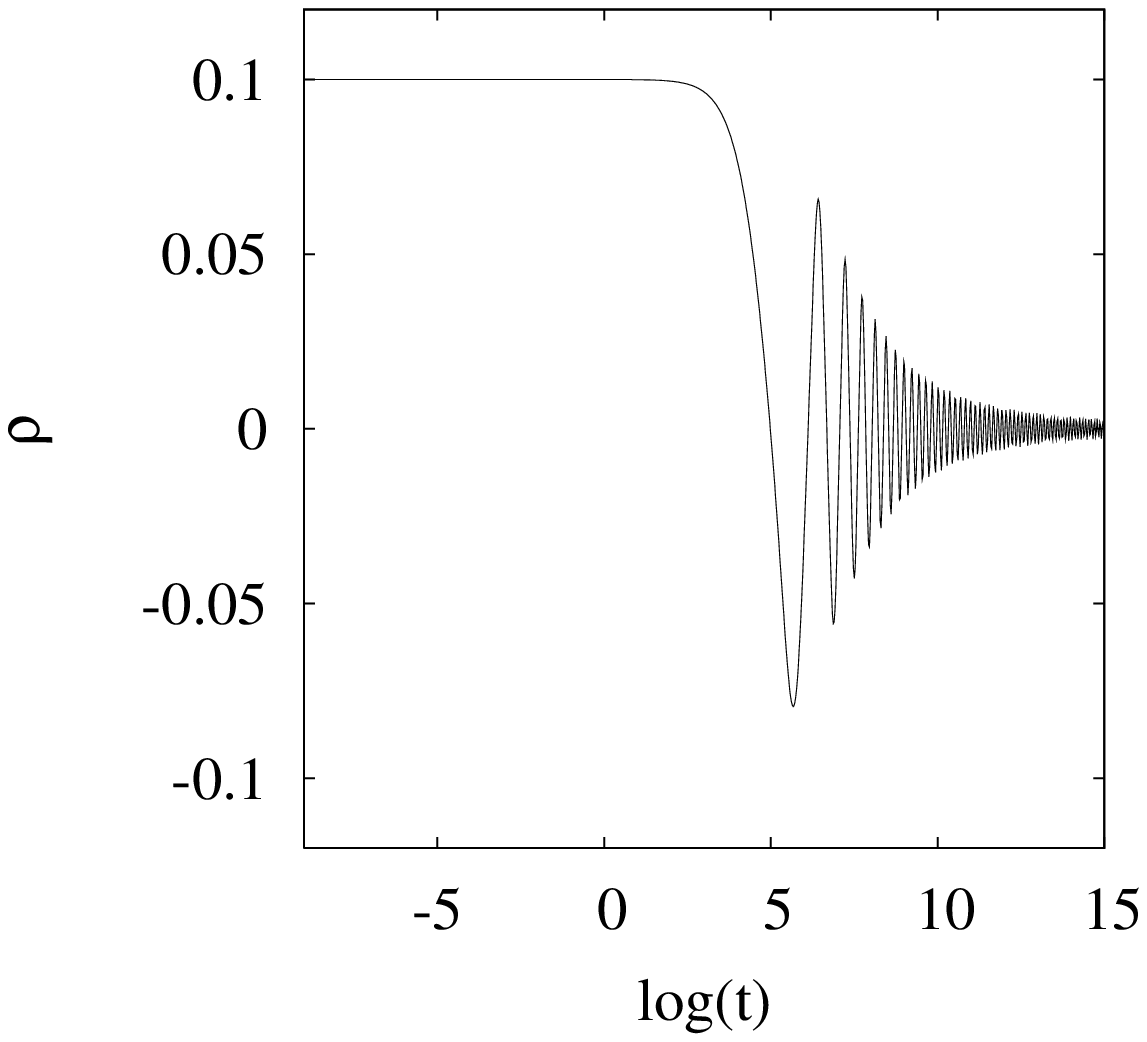}&
\includegraphics[%
  scale=0.5]{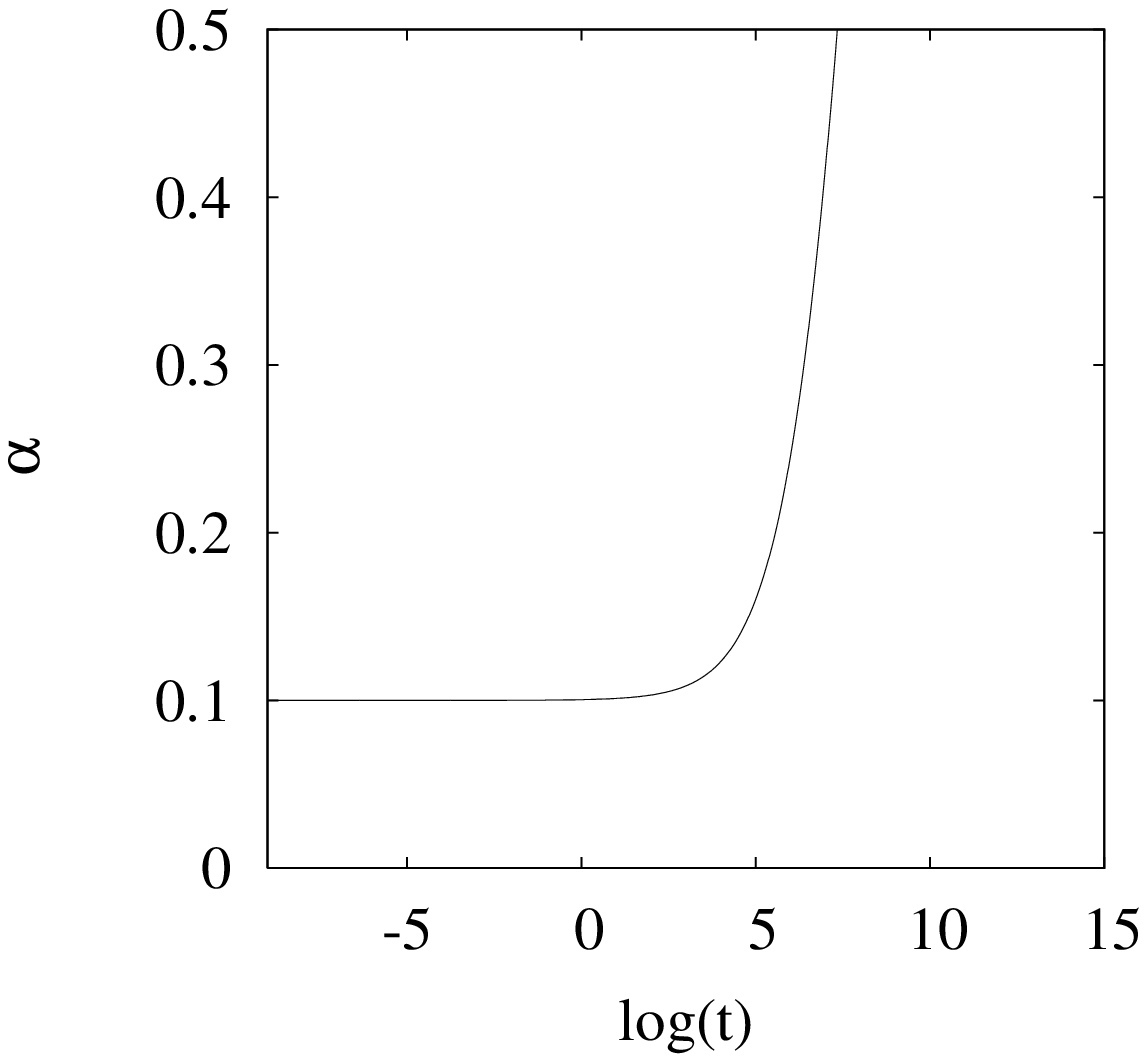}\tabularnewline
\end{tabular}\tabularnewline
\includegraphics[%
  scale=0.5]{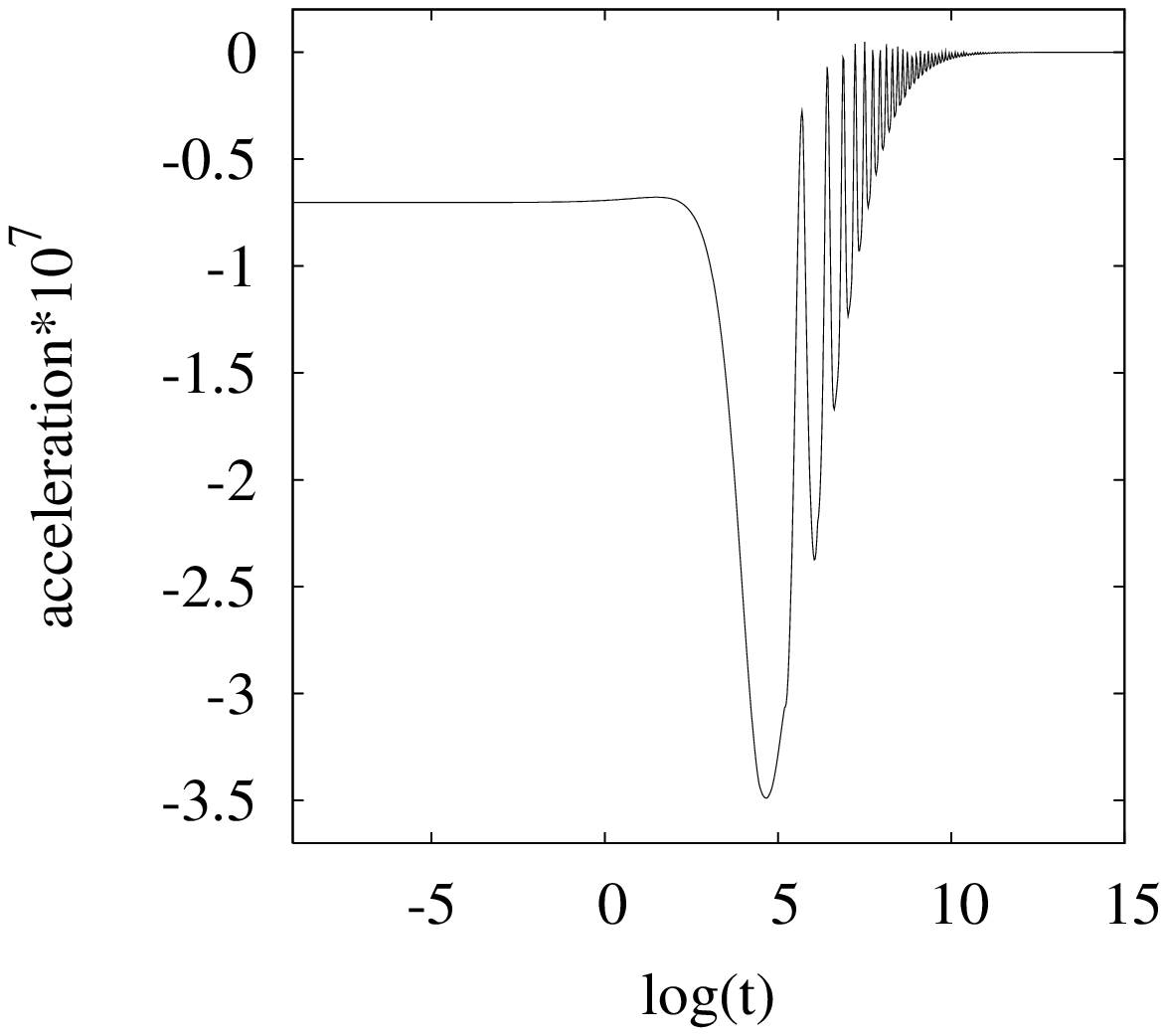}\tabularnewline
\end{tabular}\end{center}

\caption{Hubble damped oscillations of a dissolved D3-brane and the resulting
(lack of) acceleration.}
\end{figure}

Finally if we begin with non-zero $\rho$ and also slightly excite
the brane separation modulus the situation is very different. Indeed
we find the reverse of the initially separated but excited D-branes;
we would expect large enough excitations of the brane separation to
cause a branch change when the system reaches the hub of the cross
shaped potential. Such a branch change would correspond to the instanton
being emitted as a D3-brane from the D7-brane stack. As the plots in 
figure 6 clearly show, the D3-brane can indeed be emitted in this manner.

\begin{figure}
\begin{center}\begin{tabular}{c}
\begin{tabular}{cc}
\includegraphics[%
  scale=0.5]{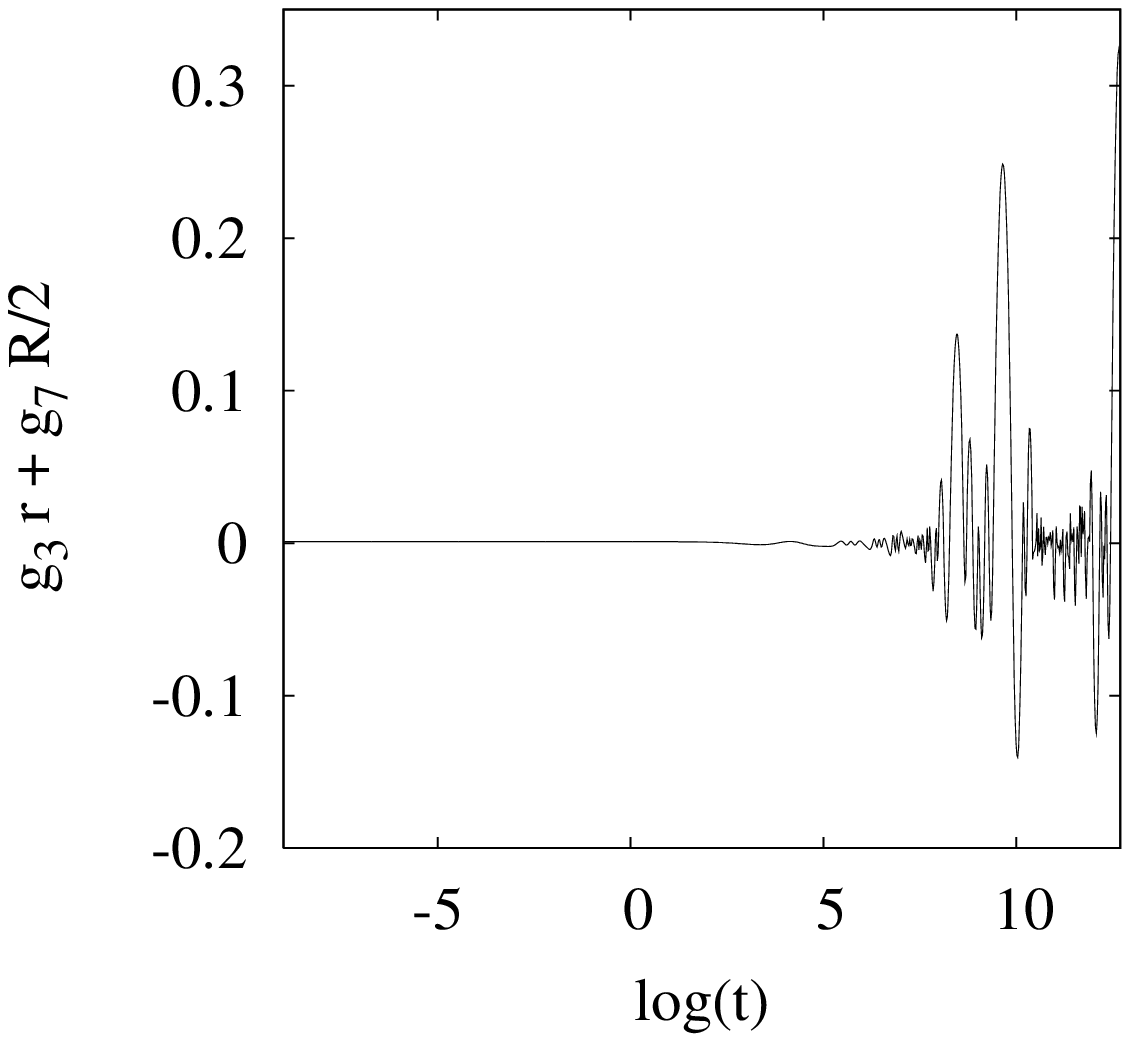}&
\includegraphics[%
  scale=0.5]{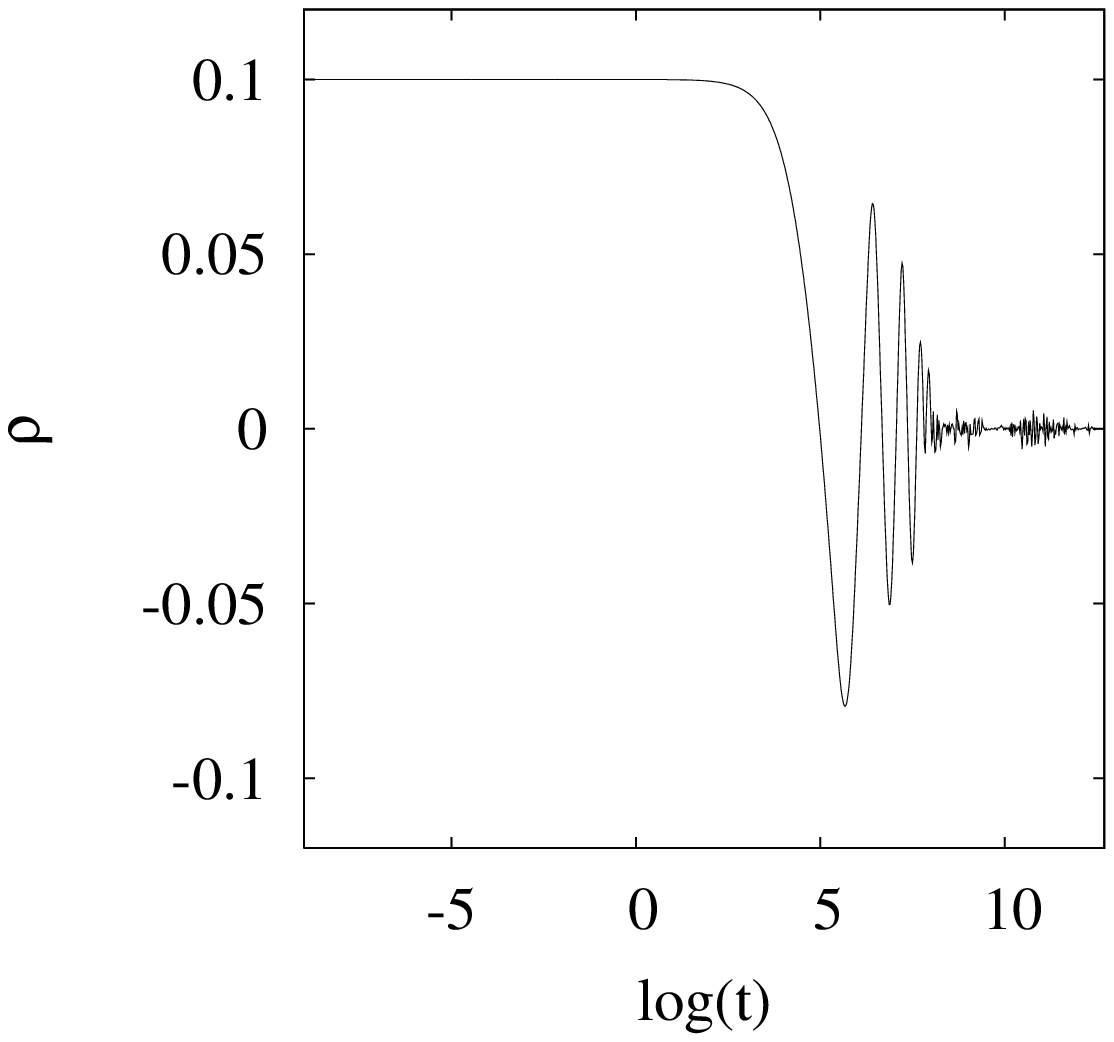}\tabularnewline
\end{tabular}\tabularnewline
\includegraphics[%
  scale=0.5]{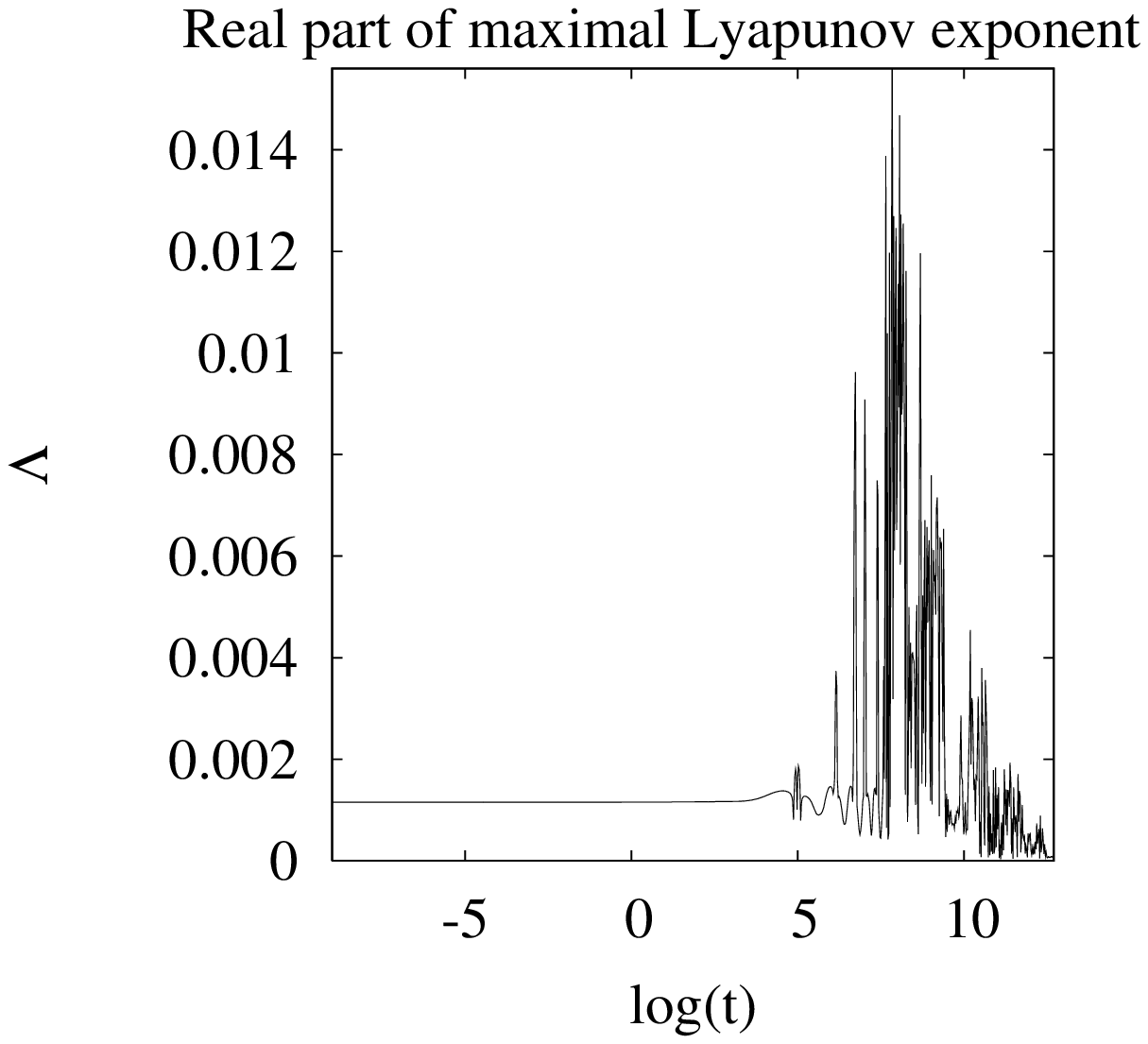}\tabularnewline
\end{tabular}\end{center}

\caption{Emission of a slightly excited dissolved D3-brane.}
\end{figure}

\subsection*{\textmd{\emph{Multiple D7-branes}}}

The previous discussion can be extended to consider multiple D7-brane
stacks. Consider beginning with a single stack of 4 D7-branes beginning
with a gauge group of $U(4)$. Keeping the previous truncation but
with the division \begin{equation}
C^{7}=\left(\begin{array}{cc}
Re^{i\psi}\frac{1}{2}\mathbf{I}_{2} & \mathbf{0}\\
\mathbf{0} & R_2 e^{i\psi_2}\frac{1}{2}\mathbf{I}_{2}\end{array}\right),\end{equation}
the four component twisted states in the ADHM truncation take the
form \begin{eqnarray}
A_{\alpha} & = & (\rho(-i\theta_{3}+\theta_{4}),\,\rho(-i\theta_{1}-\theta_{2}),\,\rho_2(-i\varphi_{3}+\varphi_{4}),\,\rho_2(-i\varphi_{1}-
\varphi_{2}))\nonumber \\
B^{\alpha} & = & (\rho(i\theta_{1}+\theta_{2}),\,\rho(-i\theta_{3}+\theta_{4}),\,\rho_2(i\varphi_{1}+\varphi_{2}),\,\rho_2(-i\varphi_{3}+\varphi_{4})).\label{adhmn2b}\end{eqnarray}
The system now describes two stacks of D7-branes whose positions in
the transverse space are given by $R,\, R_2$. When $R\neq R_2$, the
D7 gauge group is broken to $U(2)\times U(2)$, and the parameters
$\rho$ and $\rho_2$ describe two orthogonal small instanton transitions
with the D3-brane dissolving on either of the two D7-branes. With
the definitions \begin{eqnarray}
Re\{ S\} & = & e^{\phi}+\frac{1}{2}(R^{2}+R_2^{2})\nonumber \\
Re\{ T\} & = & e^{\beta}+\frac{1}{2}(\rho^{2}+\rho_2^{2})\nonumber \\
Re\{ T_{3}\} & = & e^{\beta_{3}}+\frac{1}{2}r^{2}\end{eqnarray}
the action and equations of motion are extended in the obvious way.

\begin{figure}
\begin{center}\begin{tabular}{c}
\begin{tabular}{cc}
\includegraphics[%
  scale=0.5]{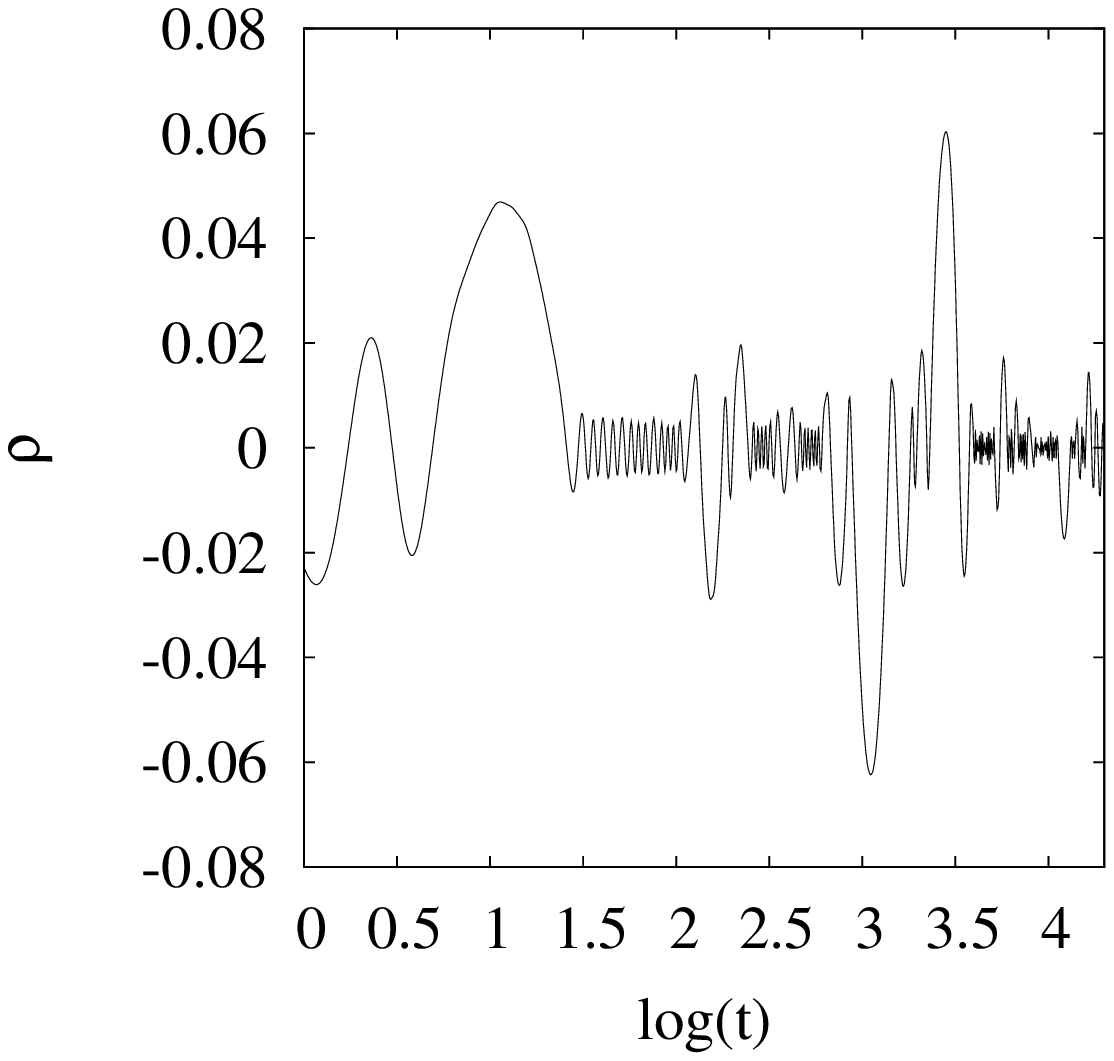}&
\includegraphics[%
  scale=0.5]{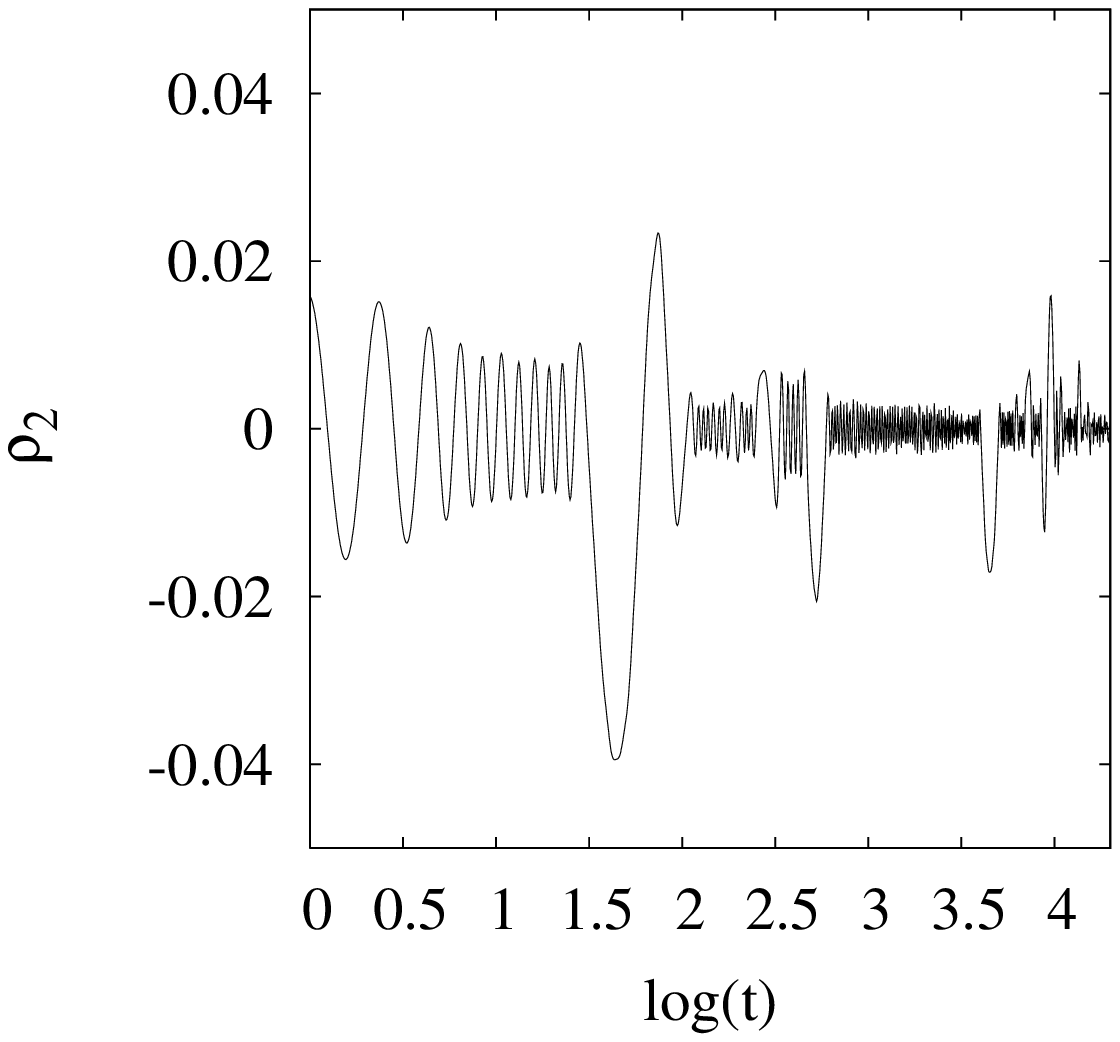}\tabularnewline
\end{tabular}\tabularnewline
\begin{tabular}{cc}
\includegraphics[%
  scale=0.5]{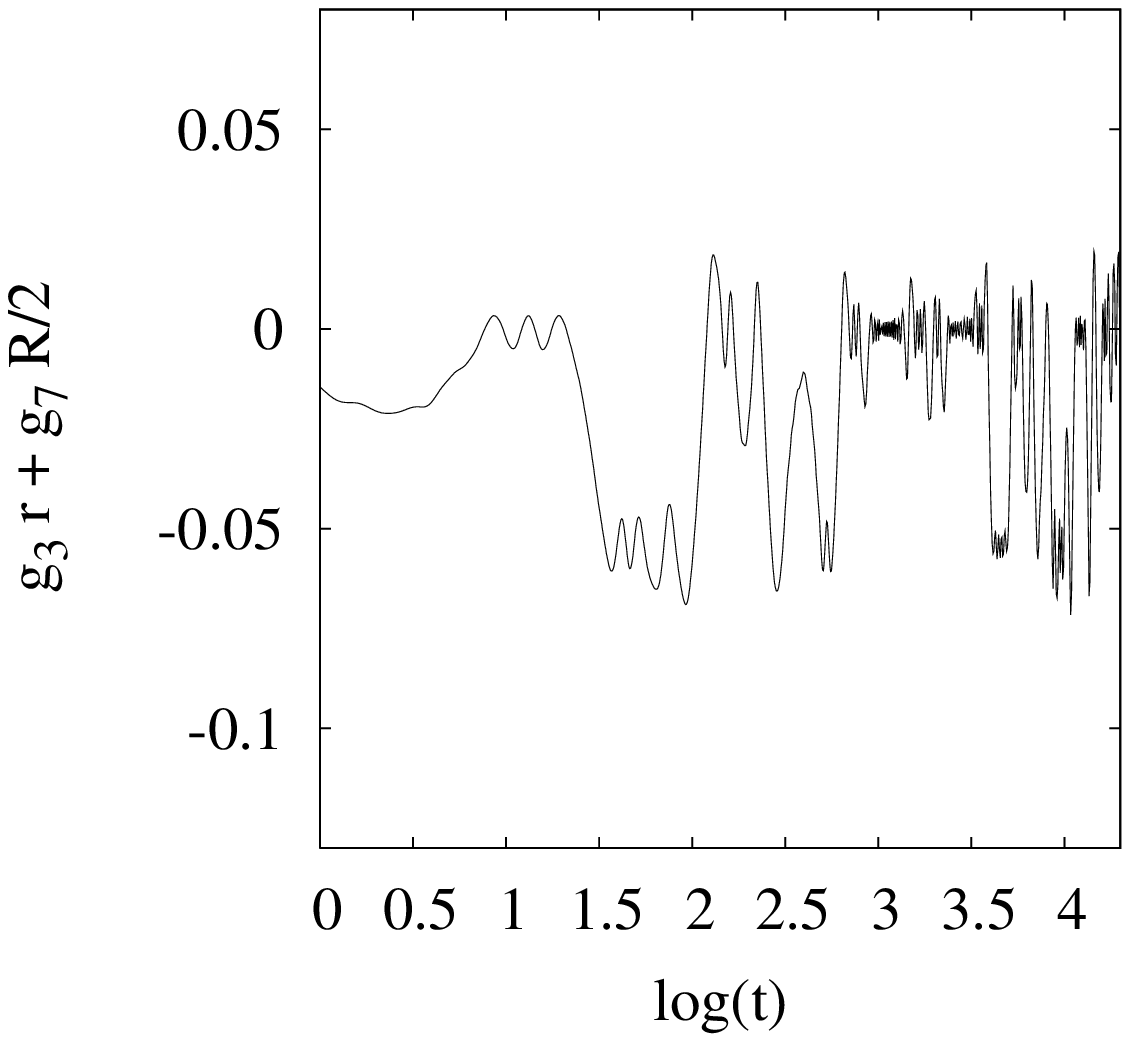}&
\includegraphics[%
  scale=0.5]{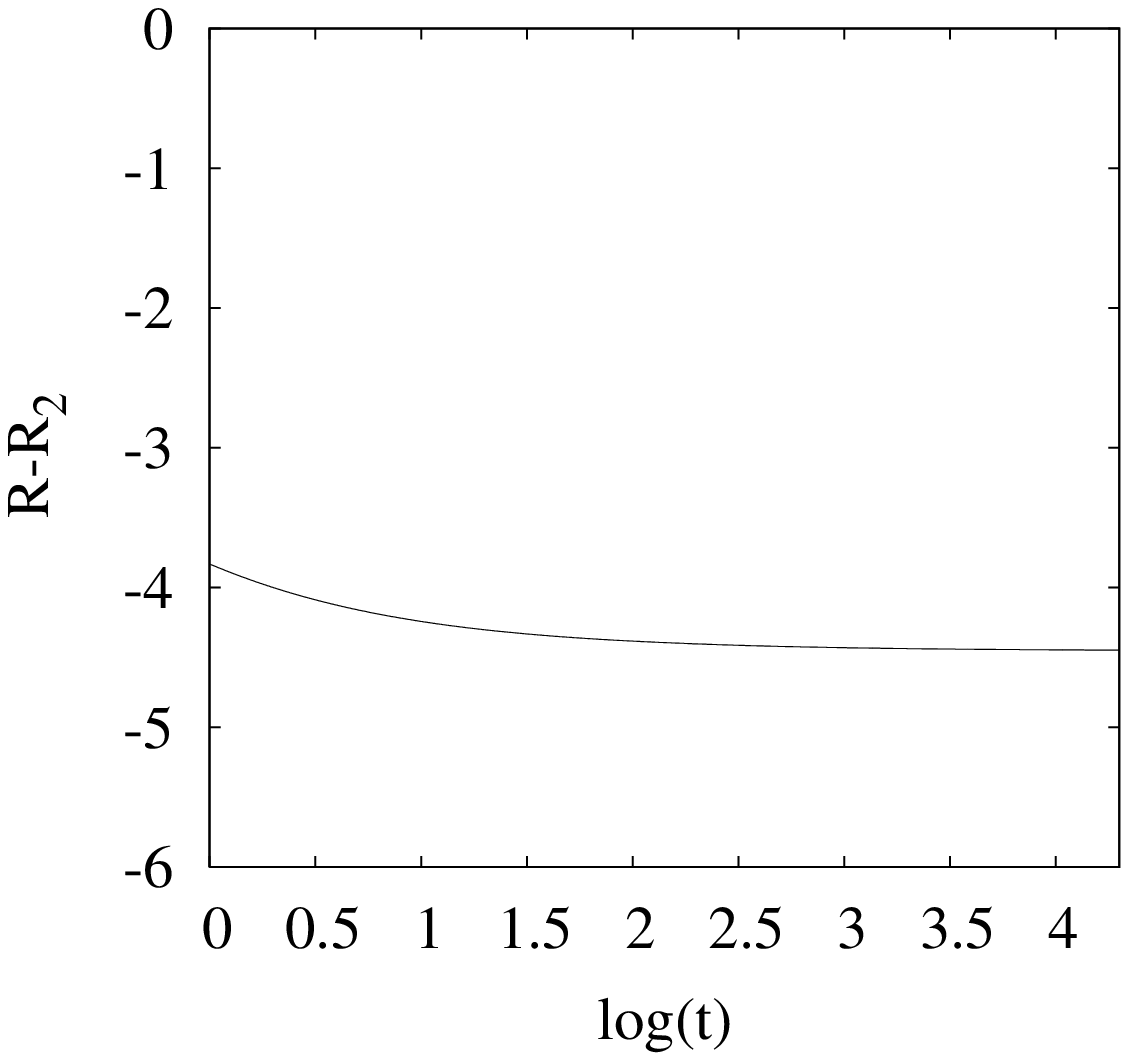}\tabularnewline
\end{tabular}\tabularnewline
\end{tabular}\end{center}

\caption{\label{cap:d3 brane swopping}D3 {}``ping-pong'': a D3 being transferred
back and forth between D7-brane stacks. }
\end{figure}

\begin{figure}
\begin{center}\begin{tabular}{c}
\begin{tabular}{cc}
\includegraphics[%
  scale=0.5]{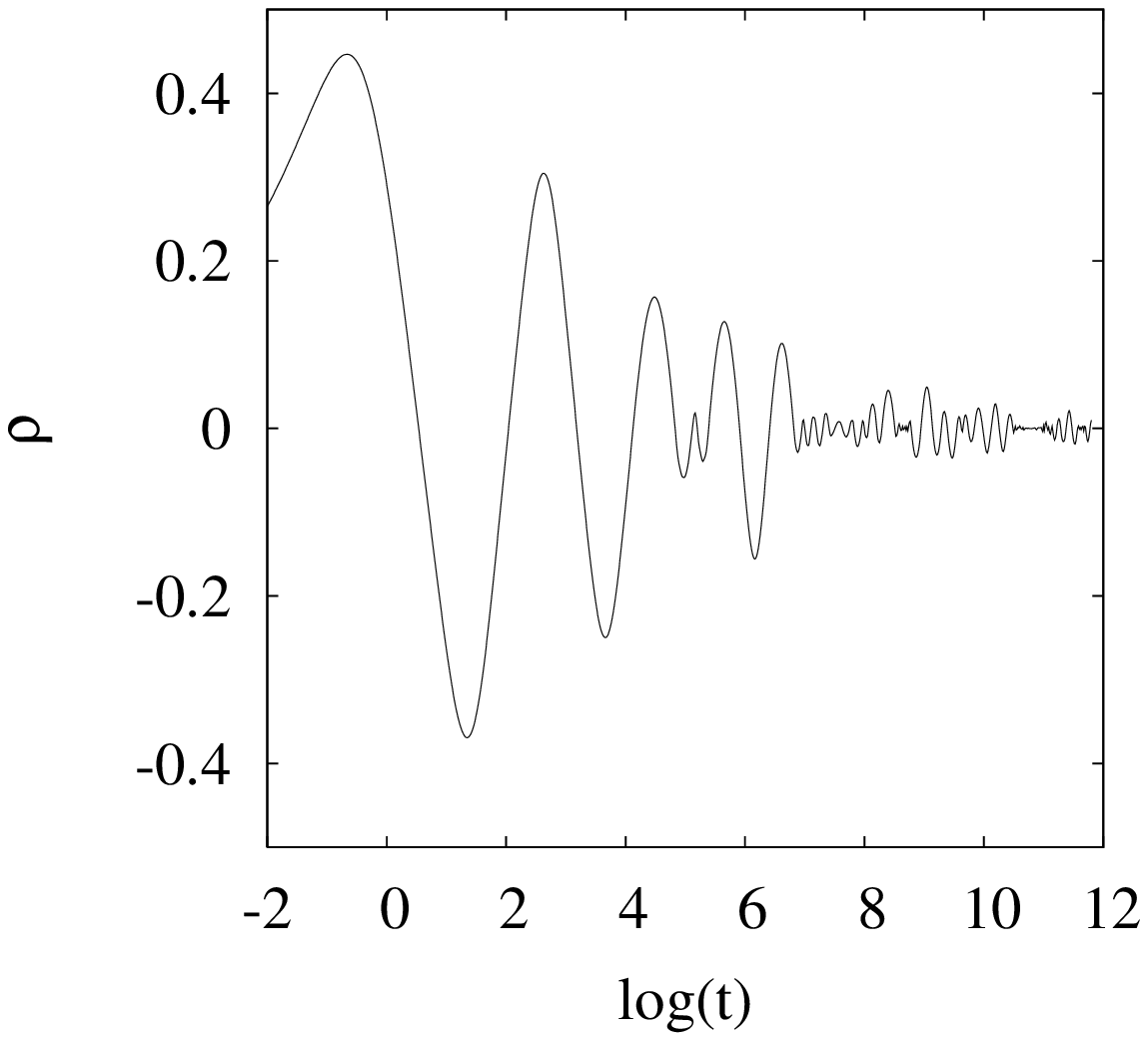}&
\includegraphics[%
  scale=0.5]{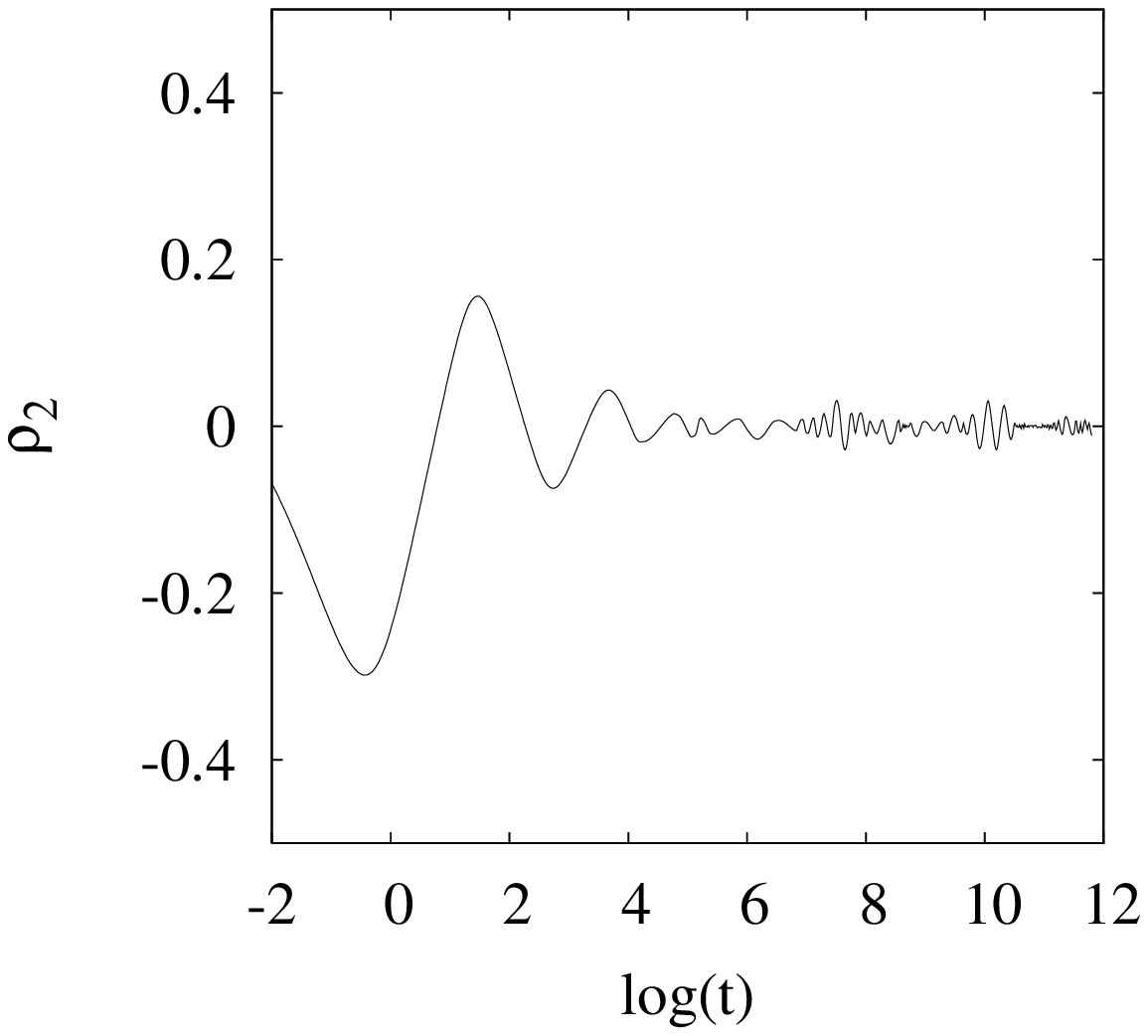}\tabularnewline
\end{tabular}\tabularnewline
\begin{tabular}{cc}
\includegraphics[%
  scale=0.5]{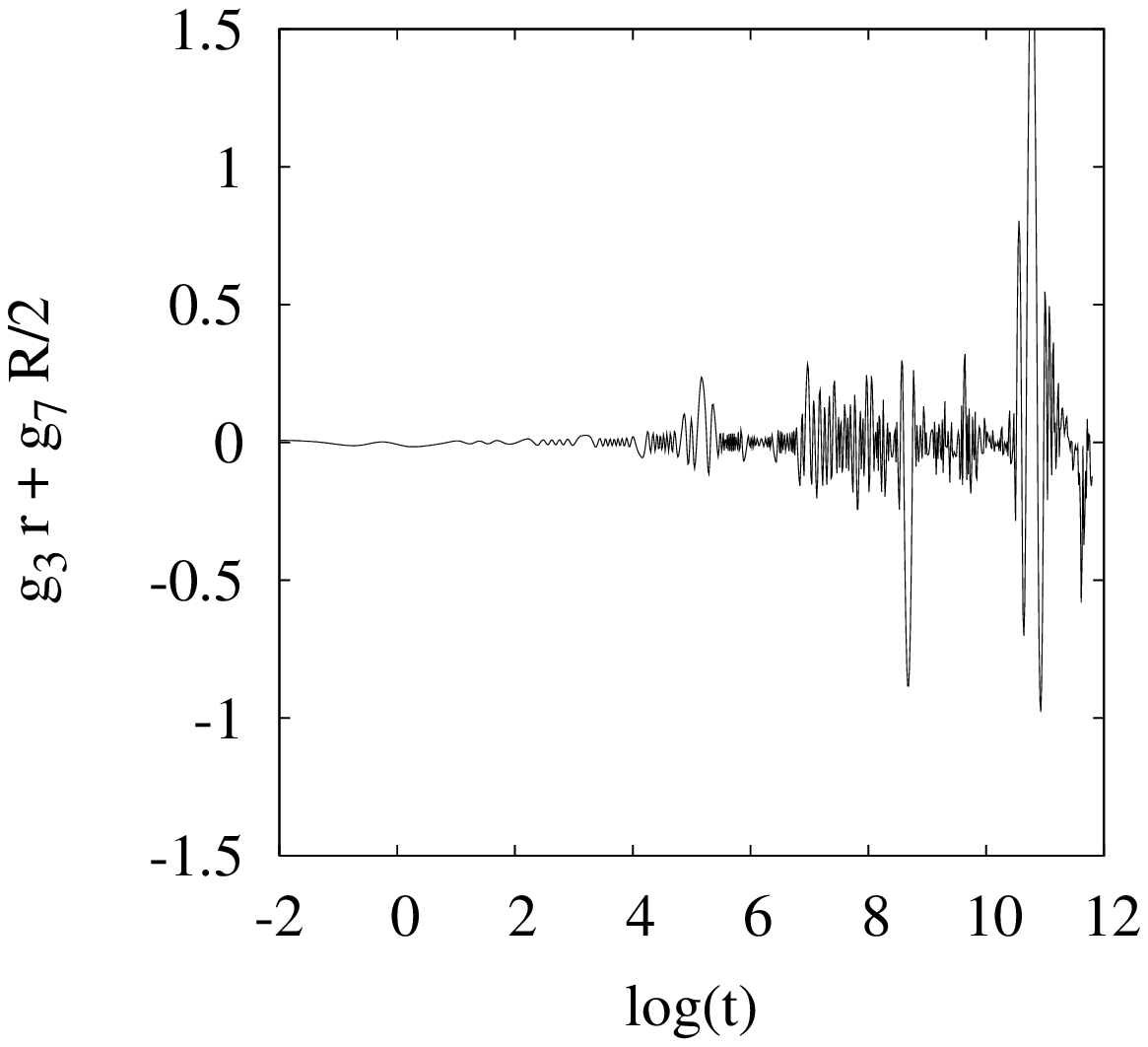}&
\includegraphics[%
  scale=0.5]{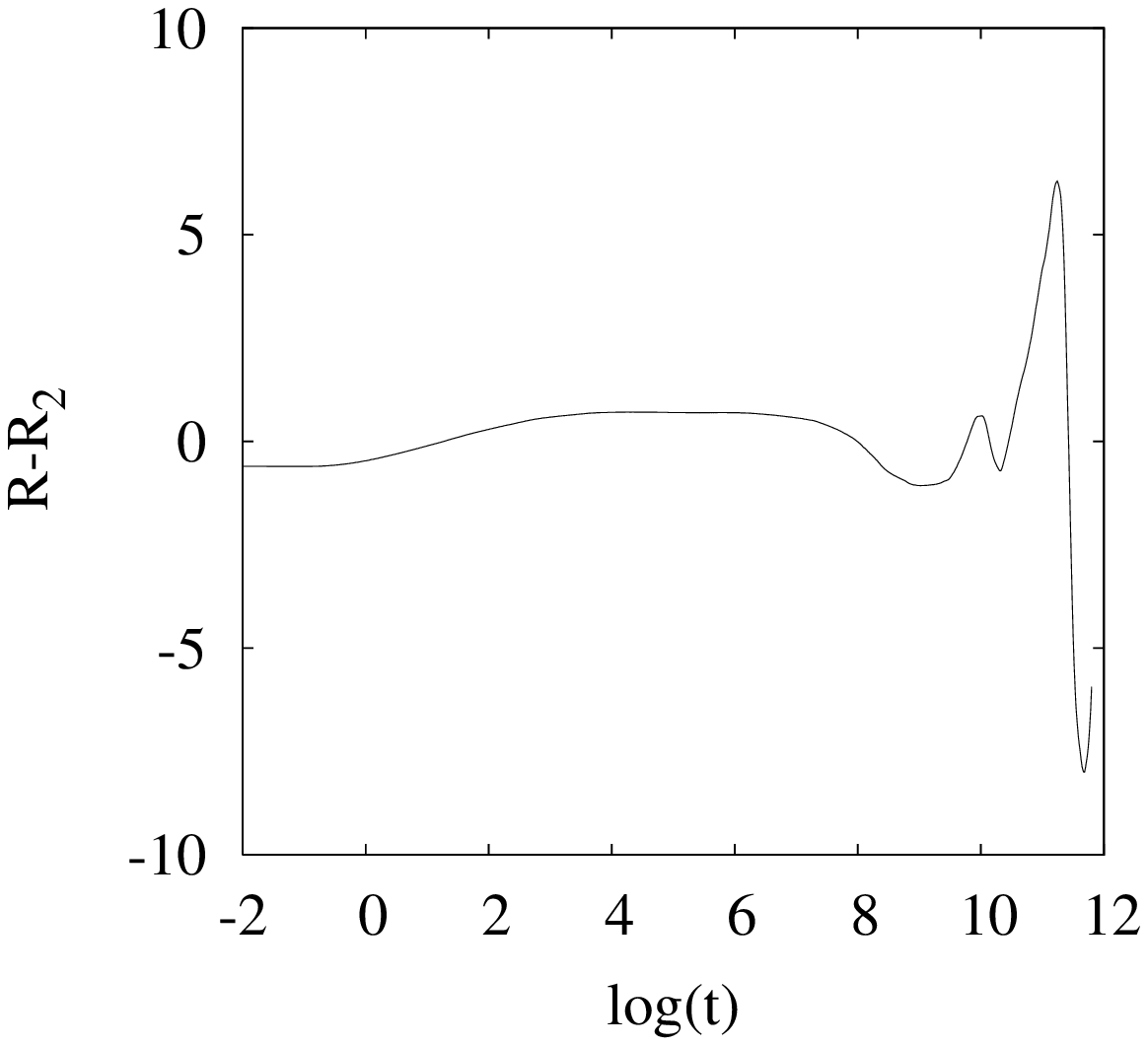}\tabularnewline
\end{tabular}\tabularnewline
\end{tabular}\end{center}

\caption{\label{cap:d7 brane attraction}Two D7-branes being attracted by
coupled excited small instantons. Note that here the D3-brane is stuck
on the D7-brane stack with position $R$ for most of the evolution, but eventually begins to oscillate freely between them.}
\end{figure}

There are two interesting types of behaviour that can be observed
in this extended system. The first (and possibly less surprising)
is that the D3-brane can be emitted from one D7-brane and captured
by the second, and indeed this can happen a number of times before
Hubble damping sets in. An example is shown in figure~\ref{cap:d3 brane swopping}. In this case (as seen in the $R-R_2$ figure) the D7-branes are, in the 
region shown, essentially undergoing the free-field evolution of figure 1,
and in fact the motion is just part of the typical `s-curve'. This is because the $g_7$ gauge coupling has deliberately been chosen to be quite small in order to decouple their motion from the D3-brane. The latter is evidently ``stuck'' on one of the D7 stacks 
for long periods while $\rho$ or $\rho_2$ are non-zero, 
before being transmitted between them on relatively 
short timescales when $\rho$ or $\rho_2$ pass through zero.
The second type of behaviour is less obvious but actually more commonplace:  
 if both $\rho$ and $\rho_2$ are slightly
excited initially, then their subsequent oscillations cause the 
D7-branes to oscillate about each other, due to the time-averaged $\rho$ and
$\rho_2$ oscillations giving an effective potential for the  
separation $R-R_2$. 
An example of this sort of behaviour is shown in 
figure~\ref{cap:d7 brane attraction} where the D7-brane displacement
goes through zero where the gauge symmetry is enhanced to $U(4).$ 
Naturally at this point the D3-branes find it easiest 
to transfer between D7-branes. Note that the timescale
for D7-brane attraction is exponentially longer than that for $\rho$
and $\rho_2$ oscillations, which in turn is exponentially longer than
the timescale for D3 oscillations. The timescales for these oscillations 
are chiefly determined by the $g_3,g_7$ couplings and the initial displacements.
Whether the D7-branes recombine
to give an enhanced $U(4)$ symmetry before the Hubble damping sets
in depends on all these factors, but the generic behaviour is clearly 
that all the fields, including the D7 stacks, oscillate around the single 
symmetric point where $R=R_2$ and all the other fields are zero.
(We consistently truncated various
light states in the description of the physics here in order to maintain
clarity, so the discussion of this point is somewhat qualitative.)

\subsection*{\textmd{\emph{Chaotic behaviour and Lyapunov exponents}}}

We have seen that small instanton transitions can indeed dynamically
occur in phenomenological string settings, but we have found that 
the collision of a D3-brane with a D7 stack does 
not \emph{per se} guarantee that such
a transition will take place and even if it does the behaviour does
not correspond to the smooth monotonic expansion of a dissolved 
D3-brane that might naively have been expected. Likewise the emission
of a D3-brane from the D7 stack is not guaranteed when an instanton
shrinks to zero size and, even if a small instanton transition does
occur, it does not typically result in the newly created extended
object escaping to infinity. 

The final state of the system is in fact strongly dependent upon the
initial conditions. In particular the degree of initial excitation
of the twisted sector and D-brane separation moduli is the dominant
factor in deciding what qualitative dynamical behaviour the system
will exhibit. If both the instanton size and D-brane separation moduli
are initially fairly excited, the system's final state tends to be
described by unpredictable chaotic oscillations about the small instanton
transition point. The physical origin of this oscillatory behaviour
is due to the potential being dominated by a term proportional to
$\rho^{2}(g_{3}r+\frac{g_{7}}{2}R)^{2}$ in the vicinity of the transition
point. Thus an expectation value for $\rho$ results in an effective
mass term for the D-brane separation and \emph{vice versa}. Small
oscillations of $\rho$ , for example, then result in a time varying
mass term for the D-brane separation modulus. The separation degree
of freedom then oscillates in this time dependent potential well which
in turn causes a contribution to the effective potential seen by the twisted
sector, and so on. 

The chaotic nature of the resulting oscillations should come as no
surprise in a non-linear system such as this. Indeed the system with
canonical kinetic terms in two fields $X$ and $Y$, with an $X^{2}Y^{2}$
potential is a well known partially chaotic system that has been studied
in some depth \cite{steeb,nip,dahl}. This system (which we refer
to as the '$XY$ system') is coincident with ours in the limit where
the $\rho^{4}$ term in our potential is negligible and in the limit
where the dynamics of the metric moduli and dilaton are associated
with timescales much longer than those governing the oscillations
we have been describing. The $XY$ system in fact exhibits chaotic
behaviour for the vast majority of its possible initial conditions.
Thus we expect that for a large class of initial conditions our system
will also exhibit chaotic oscillatory behaviour with the oscillations
tending to decay at late times due to Hubble damping. 

That our system is truly chaotic can be confirmed by calculating the
relevant Lyapunov exponents. If one of these exponents has a positive
real part when the system appears to be unpredictable then the behaviour
is truly chaotic in the sense that adjacent trajectories in phase
space diverge exponentially. Plots of the real part of the maximal
Lyapunov exponent $\Lambda_{max}$ can be found in the figures. It
should be noted that the inverse Lyapunov exponents also contain information
about timescales. For example, in figure~\ref{cap:caption-number-two}
the Lyapunov exponent is non-zero but the behaviour appears to be
quite predictable in the plot because the timescale $1/\Lambda_{max}\sim10^{4}$
is comparable in size with the total time of the evolution shown.
By contrast figure~\ref{onagainoffagain} is unpredictable because
$1/\Lambda_{max}\sim 30$ is very short on the timescale of the plot.
Note that the calculation of the Lyapunov exponents took into account
all of the fields, including the compactification moduli and the dilaton,
and so the effect of Hubble damping on the chaotic behaviour is already
included in the plots. The behaviour is usually (but not always) tending
to small Lyapunov exponents at large times. 

An additional feature of the evolutions we can clearly see is that
Hubble damping causes the {}``matter'' fields $r,\rho,R$ to sample
a surface in phase space that is shrinking as energy is Hubble damped
away. This effect is not completely obvious as it depends on how fast
the potential (which is dependent on for example $\beta$) is flattening
out, but we found no counterexamples. One should also bear in mind
though that, as is well known from previous studies 
\cite{Lidsey:1999mc,Copeland:2001zp},
the motion of the brane position moduli tends to freeze asymptotically 
(for sufficiently small excitations of the twisted fields) due to the 
cosmological evolution of $\alpha,\beta,\beta_{3}$, and it is not clear 
in any particular case which effect will dominate. 

Of course in the chaotic regimes (i.e. where the inverse Lyapunov
exponent is shorter than the timescales of interest) it is impossible
to predict when the system will be in the Higgs or Coulomb branch
(-like) phase, but at late times we see that the system tends to the
point where the maximum number of light states occurs. 
This behaviour has a passing resemblance to the attraction
to {}``extra species points'' noted in ref.\cite{Kofman:2004yc}
although, since in that case the effect was quantum mechanical whereas here
it is purely classical, it actually has more in common with the earlier 
results presented in
refs.\cite{Brandle:2002fa,Jarv:2003qy,Mohaupt:2004pr,Lukas:2004du}.
We comment further on these similarities in the discussion. 


\section{Discussion}

The behaviour we have described here is formally similar to other
systems in string cosmology, particularly those where extra light
states appear as a consequence of the evolution. A comparison can
be drawn for example with the results of 
refs.\cite{Brandle:2002fa,Jarv:2003qy,Mohaupt:2004pr,Lukas:2004du}.
The system in those cases involved flop and conifold transitions in
both string and $M$-theory compactifications. We would expect 
chaotic dynamics to be 
exhibited by these systems for appropriate ranges of initial conditions. 
Indeed indications of chaotic oscillations are evident in the plots of 
ref.\cite{Lukas:2004du}. 

As we noted in the previous section the behaviour is also qualitatively
similar to that found in ref. \cite{Kofman:2004yc}. In that work
it was argued that systems which have a point where the number of light
species is enhanced, a so-called {}``enhanced species point'' (ESP),
will experience damping due to quantum mechanical production of the
light states every time they pass through these ESPs. In our case the
ESP is the small instanton transition point, where the D3-brane is
undissolved (i.e. $\rho=0$) but lying on top of the D7-brane. In
fact this type of quantum mechanical effect \emph{will} occur in our
system as well. The states in question are precisely the open string
'twisted' states stretching from D3 to D7-brane: on each pass through
the ESP such states are produced and the energy required to stretch
them puts a brake on the motion of the D3-brane with respect to the
D7-brane \cite{Kofman:2004yc,douglasbachas}. However at weak
coupling and low velocities such effects are subdominant, with the
evolution being essentially classical. What we can conclude from the
present discussion is that \emph{even classically} the D3/D7 system
is driven to the ESP. A similar argument applies in systems with initially
separated D7-branes. In that case the system is not only driven to
the small instanton transition point, but the D7-branes are themselves
attracted to each other and hence enhanced gauge symmetry as well.
Along with this conclusion for the D3/D7 system, comes an important
implication for model building in intersecting branes; since recombination
transitions are T-dual (in a generic sense) to D3/D7 systems, it seems
that such systems, if they have a period of unconstrained cosmological
evolution, will naturally end up maximizing the number of light chiral
states and gauge symmetries consistent with the topology of the initial
configuration. 

The fact that we obtain the chaotic behaviour described above
for a very large range of initial conditions has significant implications
for the cosmology of D-brane models. For example consider what the
dynamics shown in figure~\ref{onagainoffagain} implies for the cosmology
of the early Universe in a situation where the temperature in the
Universe is negligible when compared to the energies involved in our
numerical solutions. The $\rho$ and D-brane separation spend long
periods of time being essentially unexcited followed by periods where
they oscillate considerably. This corresponds to a series of moduli
driven gauge symmetry changing phase transitions \cite{Bastero_gil:2002hs,Gray:2004rw}
in which the gauge group is in turns enhanced and then reduced in
rank. The mass scale of gauge symmetry breaking in the Higgs phase
decreases with time and eventually goes to zero as the system settles
down at the symmetric point. Such a sequence of phase transitions
has not to our knowledge been considered in cosmology, but in our analysis
we see that it is in fact quite typical; the usual picture in which
the Universe undergoes sequential reduction of gauge symmetry in a
series of phase transitions is not by any means universal in D-brane
systems.

There are in addition many discussions of inflationary models involving D-branes. 
Refs.\cite{kalloshd3d7,Dasgupta:2004dw, Shandera:2004zy} 
specifically concerned D3/D7-brane systems and invoked various mechanisms for 
giving the slow-roll potential; for example in ref.\cite{kalloshd3d7},
fluxes give a logarithmic potential in the Coulomb branch which causes
inflation. (In a dual system the potential in question can be put
down to the presence of non-zero brane angles corresponding to non-zero
Fayet-Iliopoulos terms; the resulting model is known as {}``P-term''
inflation \cite{pterm}.) The D3-brane rolls towards the D7-brane and dissolves
on it, the dissolution playing the role of the {}``waterfall'' stage
of hybrid inflation. Our discussion is relevant to the dynamics of
this type of phase transition. Indeed one interesting possibility
is an explanation for the starting configuration. For the scenario
to work one requires a sufficiently damped (i.e. with kinetic energy
negligible compared to potential energy in order to satisfy the slow-roll
conditions) D3-brane sitting out along the potential. The natural
question to ask is how it got there. Here we have seen that it is
quite natural for the D3-brane to have been emitted from a neighbouring
D7-brane. Also in this context the production of multiple stages of
inflation seems likely. Therefore it would be interesting to repeat 
the analysis for that case. 
The expectation is that once the flat directions are
lifted (to generate inflation) the system will behave much as described
in ref.\cite{easther}: chaotic behaviour will occur when the energy
is higher than the barriers in the potential. There could be various
interesting consequences such as for example enhanced production of
defects such as cosmic strings. 
This is intimately related to the presence of chaos: inflation
aims to solve the defect problem by exponentially suppressing variation
in field values, but chaos makes the evolution exponentially sensitive
to initial conditions.

\section{Conclusion}

In this paper we have described in microscopic detail the cosmological
dynamics associated with small instanton (and brane recombination) 
transitions in type II string theory for the first time. 
We have seen that the naive picture of
a D3-brane colliding with a D7 stack and becoming an instanton which
monotonically increases in size with time is not in fact correct.
Instead if such a phase transition does occur it tends to be chaotic
with unpredictable oscillations about the point of transition. 
This results in various unusual types of behaviour. For example 
fairly generic initial conditions result in 
a sequence of symmetry breaking and restoring phase
transitions, and hence a fluctuation of the low energy gauge
group. At late times the system tends to points with enhanced numbers 
of light states, where the D3-branes are stuck on the D7-branes,
as has been argued for in different contexts in the past.
We also observe that at late times the D3-branes
can sporadically jump between the different D7 stacks.

All of this behaviour could have significant implications for early 
Universe cosmology and for D-brane model building in general.
A typical supersymmetric phenomenological 
construction consists of a number of
separated D3 and D7-branes. Intuitively one imagines 
that, before all the moduli are (presumably) stabilised, the
D-branes smoothly ``glide about'' in near BPS configurations;
the present analysis shows that their behaviour is in fact 
more interesting and rather more complicated.

\bigskip{}
\noindent \textbf{\large Acknowledgements}{\large \par}

\medskip{}
\noindent We thank D.Tong and V.V.Khoze for useful discussions.
J. G. is supported by PPARC.

\newpage

\section{Appendix A: Equations of Motion}

We present in this appendix the equations of motion which follow from
the action in eq.(\ref{eq:action}) and the metric ansatz $ds^{2}=-dt^{2}+e^{2\alpha}d\vec{x}^{2}$.
All fields and the metric function $\alpha$ are taken to be functions
of $t$ only and derivatives with respect to this coordinate are denoted
by a prime.

\begin{eqnarray*}
(12\alpha'e^{3\alpha})' & = & 3e^{3\alpha}\left[\frac{3}{16}g_{7}^{2}\rho^{4}e^{-2\beta-\beta_{3}}+\rho^{2}(g_{3}r+g_{7}\frac{R}{2})^{2}e^{-\phi-\beta-\beta_{3}}\right]\\
(-e^{3\alpha}\beta')' & = & e^{3\alpha}\left[\frac{1}{2}e^{-\beta}\rho'^{2}-\frac{3}{16}g_{7}^{2}\rho^{4}e^{-2\beta-\beta_{3}}-\frac{1}{2}\rho^{2}(g_{3}r+g_{7}\frac{R}{2})^{2}e^{-\phi-\beta-\beta_{3}}\right]\\
(-e^{3\alpha}\frac{\phi'}{2})' & = & e^{3\alpha}\left[\frac{1}{2}e^{-\phi}R'^{2}-\frac{1}{2}\rho^{2}(g_{3}r+g_{7}\frac{R}{2})^{2}e^{-\phi-\beta-\beta_{3}}\right]\\
(-e^{3\alpha}\frac{\beta_{3}'}{2})' & = & e^{3\alpha}\left[\frac{1}{2}e^{-\beta_{3}}r'^{2}-\frac{3}{32}g_{7}^{2}\rho^{4}e^{-2\beta-\beta_{3}}-\frac{1}{2}\rho^{2}(g_{3}r+g_{7}\frac{R}{2})^{2}e^{-\phi-\beta-\beta_{3}}\right]\\
(-e^{3\alpha-\beta_{3}}r')' & = & e^{3\alpha}\left[g_{3}(g_{3}r+g_{7}\frac{R}{2})\rho^{2}e^{-\phi-\beta-\beta_{3}}\right]\\
(-e^{3\alpha-\phi}R')' & = & e^{3\alpha}\left[\frac{g_{7}}{2}(g_{3}r+g_{7}\frac{R}{2})\rho^{2}e^{-\phi-\beta-\beta_{3}}\right]\\
(-e^{3\alpha-\beta}\rho')' & = & e^{3\alpha}\left[\frac{3}{8}g_{7}^{2}\rho^{3}e^{-2\beta-\beta_{3}}+(g_{3}r+g_{7}\frac{R}{2})^{2}\rho e^{-\phi-\beta-\beta_{3}}\right].\end{eqnarray*}
In addition we obtain the following Hamiltonian constraint for the
system:

\[
6\alpha'^{2}=\frac{1}{4}\phi'^{2}+\frac{1}{2}e^{-\phi}R'^{2}+\frac{1}{4}\beta_{3}'^{2}+\frac{1}{2}\beta'^{2}+\frac{1}{2}e^{-\beta}\rho'^{2}+\frac{3}{32}g_{7}^{2}\rho^{4}e^{-2\beta-\beta_{3}}+(g_{3}r+g_{7}\frac{R}{2})^{2}\frac{1}{2}\rho^{2}e^{-\phi-\beta-\beta_{3}}\,.\]
It is these equations that are solved in the text, analytically for
$\rho=0$ and numerically in other cases.

\clearpage \nocite{*}
\bibliography{ag5}
\bibliographystyle{plain}

\end{document}